\documentclass[12pt]{article}

\def\be{\begin{equation}}
\def\ee{\end{equation}}
\def\bea{\begin{eqnarray}}
\def\eea{\end{eqnarray}}

\usepackage{amssymb,latexsym}
\usepackage{graphicx}
\usepackage{amssymb, amsmath, amsopn, amsthm} 
\usepackage[section]{placeins}

\makeatletter \@addtoreset{equation}{section} 
\makeatother

\begin{document}
\begin{titlepage}
\thispagestyle{empty} 
\begin{flushright}
    \hfill{CERN-PH-TH/2006-103 } \\
    SU-ITP-2006-17\\
    \hfill{hep-th/0605266}\\ 
    May 29, 2006
\end{flushright}

\begin{center}
    { \LARGE{\bf Domain walls, near-BPS bubbles\\[4mm]
and probabilities in the landscape}}
    
    \vspace{30pt}
    
  {\large  {\bf Anna Ceresole$^\star$, \ Gianguido Dall'Agata$^\dagger$, \ Alexander Giryavets$^\S$,\\[4mm]
    Renata Kallosh$^\S$ and Andrei Linde$^\S$ }}
    
    \vspace{15pt}
    
    {\it $\star$ INFN $\&$ Dipartimento di Fisica Teorica\\
    Universita' di Torino, Via Pietro Giuria 1, 10125 Torino, Italy}
    
    \vspace{10pt}
    
    {\it $\dagger$ Physics Department, Theory Unit, CERN, \\
    CH 1211, Geneva 23, Switzerland}
    
    \vspace{10pt} {\it \S \ Department of Physics,\\
    Stanford University, Stanford, CA 94305}
    
    \vspace{20pt}
 \end{center}

\begin{abstract}
    
    We develop a theory of static BPS domain walls in stringy landscape and present a large family of  BPS  walls interpolating between different supersymmetric vacua. Examples include  KKLT models, STU models, type IIB multiple flux vacua, and models with several Minkowski and AdS vacua.  After the uplifting, some of the vacua become dS, whereas some others remain AdS. The near-BPS walls separating these vacua may be seen as bubble walls in the theory of vacuum decay. As an outcome of our investigation of the BPS walls, we found that the decay rate of dS vacua to a collapsing space with a negative vacuum energy can be quite large. The parts of space that experience a decay to a collapsing space, or to a Minkowski vacuum, never return back to dS space.  The channels of irreversible vacuum decay serve as sinks for the probability flow. The existence of such sinks is a distinguishing feature of the  landscape. We show that it strongly affects the probability distributions in string cosmology.

\end{abstract}

\vspace{10pt} 
\end{titlepage}

\tableofcontents

\newpage   \baselineskip 5.5 mm

\section{Introduction}

Soon after the invention of inflationary cosmology it was realized that inflation may divide our universe into many exponentially large domains corresponding to different metastable vacuum states \cite{linde1982}.
The total number of such vacuum states in string theory can be enormously large \cite{Lerche:1986cx,Bousso:2000xa,Douglas}.
A combination of these two facts with the KKLT mechanism of vacuum stabilization \cite{Kachru:2003aw} gave rise to what is now called the string landscape scenario \cite{Susskind:2003kw}.

The purpose of this paper is to clarify some features of the landscape.
In particular, one may wonder what are the properties of the domain walls separating different vacua, how these domains may form, and how large will be the fraction of the volume of the universe inside different domains.

To this end,  we will construct a broad class of static BPS domain
wall solutions interpolating between different AdS or Minkowski
vacua with unbroken supersymmetry. Quite generally, these BPS domain walls  may interpolate between various minima, maxima, or saddle points of the scalar field
potential. 

BPS domain wall solutions in 4-dimensional $N=1$ supergravity were
first studied in \cite{Cvetic:1996vr}. The methods used for that
analysis have been later refined and generalized to a variety of
models, for different number of dimensions and supersymmetries,
see e.g. \cite{Ceresole:2001wi,Behrndt:2001mx,Louis:2006wq}. Our
present aim is to  further extend these methods to make them
suitable for investigation of the string landscape scenario.
Following the logic of the recent developments, we will begin with
an investigation of the supersymmetric  vacua arising from flux
compactifications prior to the KKLT uplifting, and later
introduce the more physically relevant
metastable de Sitter vacua.

We start in Section 2 with an  overview of some known features of
domain walls in effective d=4, N=1 supergravity. The new results
here include the derivation of BPS domain wall gradient flow
equations in generic models with F-term potentials with arbitrary
K\"ahler potential, superpotential and number of moduli. We
rewrite the gravitational action, including the boundary $K$-terms
in a BPS form: when the flow equations are satisfied, the
integrand of the action defines the BPS domain wall tension.  Afterward, in the spirit of the analogous work in five dimensions
in the context of the AdS/CFT correspondence, where domain walls
are viewed as dual to  renormalization group flows
\cite{Freedman:1999gp}, we  present the BPS flow equations in
the form of RG  flows and provide the relevant ``c-theorem''.

In Section 3 we give examples of BPS domain walls where the potential has more than one critical point but there is no barrier between them.
Our examples include an AdS minimum and an asymptotically Minkowski vacuum.
Other examples have critical points at finite distance in the moduli space: an AdS saddle point and an AdS minimum, or an AdS maximum and an AdS saddle point.
A particularly interesting case of flux vacua area codes in a Calabi--Yau hypersurface required the study of the flow of four moduli.
They flow from one critical point of the potential, the Landau--Ginsburg fixed point, towards the second critical point with larger absolute value of the cosmological constant near the conifold.

In Section 4 we study a class of models in stringy landscape with a barrier between critical points.
We have examples of a local Minkowski or AdS minimum and another absolute AdS minimum of larger absolute value of the cosmological constant.
In case of AdS-AdS wall, we display the domain walls with critical and super-critical tension.
The last case has the superpotential crossing zero between two AdS vacua.

The results of our investigation in Sections 2 - 4 directly apply to the transitions between supersymmetric AdS and Minkowski vacua.
However, the most interesting parts of the string theory landscape correspond to dS vacua.
These vacua can be obtained from supersymmetric AdS or Minkowski vacua by uplifting, which can be achieved, e.g., by adding some anti-D3 branes \cite{Kachru:2003aw}.
This procedure breaks supersymmetry and may destabilize some of the vacua, in particular those which corresponded to maxima and saddle points.
As a result, some of the BPS wall solutions may disappear after the uplifting, whereas some others will remain.
However, instead of describing the static BPS domain walls, they will describe domain walls moving with acceleration.
Such domain walls can be formed during the tunneling and bubble formation in the process of decay of metastable vacua, as discussed by Coleman and de Luccia (CDL) in \cite{Coleman:1980aw}.
We will discuss this process in Section 5.

If the uplifting is relatively small, which may happen when we uplift a supersymmetric Minkowski vacuum, or a shallow AdS vacuum, as in the version of the KKLT scenario developed in Ref.~\cite{Kallosh:2004yh}, then the properties of the domain walls do not differ much form the properties of the BPS walls prior to the uplifting.
In this case we will be talking about near-BPS, or near-extremal walls and bubbles.
The concept of near-BPS walls and bubbles is closely related to the concept of near-extremal black holes.
Whereas at the extremal limit the BPS black holes are absolutely stable and have zero temperature, the near-extremal black holes have a non-vanishing temperature, they can evaporate and eventually approach a stable extremal limit.
The physics of near-extremal black holes has been studied intensively over the years: a remarkable agreement with the physics of D-branes was established under condition of a small deviation from extremality \cite{Horowitz:1996fn}.

In our new endeavor we will try to understand the rate of tunneling into a collapsing vacuum with a negative cosmological constant from uplifted slightly non-BPS vacua using the proximity of the large size bubbles to the BPS domain walls whose tension we can evaluate exactly.
In the BPS limit the corresponding bubbles have an infinite radius, but the near-BPS ones have a large but finite radius \cite{Cvetic:1996vr}.
The description of these bubbles can be performed using the deviation from the BPS limit as an expansion parameter.
This simplifies investigation of bubble formation and vacuum decay in such models.
In Section 5 we will find, in particular, that the vacuum decay rate of the uplifted supersymmetric Minkowski vacuum does not depend on the details of the model and is given by the universal expression $e^{-{\bf S}/2}$, where ${\bf S} = {24\pi^{2}\over V_{dS}}$ is the entropy of the dS state.
This decay rate is much higher than the typical rate of decay of a dS vacuum to a Minkowski vacuum, which in many cases does not differ much from $e^{-{\bf S}}$ \cite{Kachru:2003aw}.
In application to our vacuum with $V_{dS} \sim 10^{{-120}}$, we are talking about the difference in the decay rate by a factor $\sim \exp({ 10^{{120}}/2})$.

The situation is especially interesting in a more general case when we uplift the models with several different AdS minima of the moduli potential. The AdS vacuum with a smaller value of the volume modulus experiences a bigger uplifting. The scale of supersymmetry breaking in the uplifted vacuum is related to the magnitude of the uplifting. As a result, the decay of the uplifted AdS vacuum to the collapsing space with a negative vacuum energy usually occurs much faster than the typical dS rate of the uplifted Minkowski vacuum $\sim e^{-{\bf S}/2}$.
Under some conditions, the decay rate can be quite large, with the suppression factor $e^{-{\bf S}} = \exp\bigl(-{24\pi^{2}\over V_{dS}}\bigr) \sim e^{-10^{120}}$ being replaced by $\exp\bigl(-{24\pi^{2}C\over |V_{AdS}|}\bigr) $. Here $C = O(1)$ and $|V_{AdS}|$ is the depth of the AdS minimum prior to the uplifting, which may be a hundred orders of magnitude greater than $V_{dS} \sim 10^{{-120}}$. In the KKLT-type models, the depth of the AdS vacuum prior to the uplifting is of the same order as $m^{2}_{3/2}$, where $m_{3/2}$ is the gravitino mass after the uplifting \cite{Kallosh:2004yh}. Therefore in the class of models  studied  in Section 5, the vacuum decay rate due to the tunneling to a collapsing universe  is related to the gravitino mass, $P \sim \exp \bigl(-{O(m_{3/2}^{-2})}\bigr)$. After the tunneling,  the universe within each bubble with negative vacuum energy collapses within a microscopically small time $\sim |V_{AdS}|^{-1/2}$. The condition that the lifetime of the uplifted vacuum is greater than the age of our part of the universe may serve as a vacuum superselection rule.

An important feature of the decay of dS space to a Minkowski vacuum or to a collapsing universe with a negative vacuum energy density is that the part of space that experiences such a decay never returns back to dS space.
This is quite different from the back-and-forth transitions between various dS vacua, which may keep the system in a kind of thermal equilibrium.
In this respect, the channels of irreversible decay of an uplifted dS vacuum to a Minkowski vacuum or to a collapsing universe serve as sinks for the probability flow in the landscape.
In Section 6 we will show that the existence of such sinks has important implications for the calculations of the probabilities in the landscape.

In particular, if one ignores the different rate of expansion of the universe in different vacua, i.e. uses a comoving volume distribution, then in the absence of the sinks, the fraction of the comoving volume of the universe in the state with a given vacuum energy density $V > 0$ asymptotically approaches $e^{\bf S}\sim e^{24\pi^{2}\over V}$. This distribution has a sharp maximum at the smallest possible $V$.
Meanwhile if the vacuum with the smallest possible $V$ is not the ground state but a sink, and the probability of tunneling to the sink is sufficiently large, then the fraction of the volume in the vacuum with the smallest $V$ will be exponentially suppressed, and the maximum of the probability distribution will be at  larger values of $V$.

Walls and bubbles considered in our paper interpolate between domains with different values of the scalar fields, but with the same values of fluxes.
In Section 7 we will briefly review our results and discuss a possibility of their generalization for the transitions between vacua with different fluxes.

\vskip 1 cm

\section{Supersymmetric Domain Walls in $N=1$ supergravity}

\subsection{Preliminaries}

In this paper we are going to analyze the domain wall solutions interpolating between different supersymmetric vacua of the landscape of string theory flux compactifications to 4 dimensions.
We will focus on the case of $N=1$ supergravity.
BPS domain wall solutions with 1/2 of unbroken $N=1$ supersymmetry in the minimal 4-dimensional supergravity have been discussed in \cite{Cvetic:1996vr}.
The equations presented there refer to the case of a single scalar field.
Here we are going to extend this analysis to a more general setup allowing for the possibility of many moduli fields with generic metric in the moduli space.
Of course, in any sufficiently small patch, there is a possible reparametrization of the moduli space so that there is a single scalar field supporting the domain wall.
In a rather generic setup, however, this is not possible globally, and therefore the equations we will propose in the following are more useful in this respect.
The analysis presented here is mainly based on the approach presented in \cite{Ceresole:2001wi} for the domain wall solutions of the minimal 5-dimensional supergravity.

We briefly recall the main ingredients needed in order to discuss supersymmetric bosonic configurations of N=1 supergravity coupled to chiral multiplets.
The bosonic sector of the action is 
\begin{equation}
S = \int d^{4}x \,\sqrt{-g}\left(\frac12 R - g_{i \bar \jmath} 
\partial_{\mu}z^i 
\partial^\mu \bar z^{\bar \jmath} - V(z, \bar z)\right)\,, \label{action} 
\end{equation}
where the potential can be expressed in terms of a superpotential $W(z)$ as 
\begin{equation}
V={\rm e}^{K}(g^{i\bar \jmath}D_i W \overline{D_j W} - 3 |W|^2), \label{potential} 
\end{equation}
or, in terms of the covariantly holomorphic superpotential ${\cal Z} = {\rm e}^{K/2} W$, as 
\begin{equation}
V=|D_i {\cal Z}|^{2} - 3 |{\cal Z}|^2.
\label{potential2} 
\end{equation}
The covariant derivative is defined as $D_i = 
\partial_i + \partial_iK$ on holomorphic quantities (like $W$) and as $D_i = 
\partial_i + \frac12 \partial_i K$ on covariantly holomorphic quantities (like ${\cal Z}$).
We will not consider vector multiplets and possible D-term extensions here.

The relevant supersymmetry transformations (of the gravitino and modulini) are 
\begin{eqnarray}
\delta \psi_{\mu L} & = & \left( 
\partial_{\mu}+\frac14 \omega_{\mu}^{ab} \gamma_{ab} + \frac{i}{2}A_\mu^B\right) \epsilon_L +\frac12 {\rm e}^{K/2} W \gamma_{\mu} \epsilon_{R}\,, \label{deltapsi} \\[2mm]
\delta\chi^{i}_{L} & = & \frac12 \gamma^{\mu} 
\partial_{\mu} z^{i}\epsilon_{R} - \frac12 g^{i\bar\jmath}{\rm e}^{K/2} \overline{D_{j}W}\epsilon_{L}, \label{deltachi} 
\end{eqnarray}
where $A_\mu^B \equiv \frac{i}{2}\left( \partial_\mu \bar z^{\bar \imath} 
\partial_{\bar \imath} K-
\partial_\mu z^i 
\partial_i K  
 \right)$ is the $U(1)$ K\"{a}hler connection.

\subsection{The BPS equations}\label{sec:bpsI}

In the following we will consider flat\footnote{Curved domain wall solutions have been considered for instance in \cite{Behrndt:2001mx,Freedman:2003ax}.}  space-like domain wall solutions of the supersymmetry conditions and of the supergravity equations of motion.
For this reason we are going to use the following metric Ansatz 
\begin{equation}
ds^2= a^2(r)(-dt^2 + dx^2+dy^2) + dr^2\,.
\label{typeI} 
\end{equation}
Given the domain wall structure, we split the indices $\mu = \{i,r\}$, $a= \{\underline{i},\underline{r}\}$.
For the metric (\ref{typeI}) there is only one non-trivial component of the spin connection 
\begin{equation}
\omega_j^{\underline{ir}} = a^\prime \delta_j^i, \label{connI} 
\end{equation}
where $a^{\prime} = \frac{d}{dr} a$, leading also to the Ricci scalar (which will become useful later) 
\begin{equation}
R = - 6\left[\left(\frac{a^\prime}{a}\right)^2+\frac{a^{\prime\prime}}{a}\right].
\label{scalarI} 
\end{equation}
Plugging (\ref{typeI}) into the supersymmetry equations (\ref{deltapsi})--(\ref{deltachi}) and assuming that also the scalars $z^{i}$ depend only on the radial coordinate $r$, one obtains 
\begin{eqnarray}
\delta\psi_{r L} & = & 
\partial_r \epsilon_L +\frac{i}{2}A_r^{B} \epsilon_L+ \frac12 {\rm e}^{K/2} W \gamma_r \epsilon_R \label{delpsirI} \\[2mm]
\delta \psi_{i L} & = & \frac{a^\prime}{2} \gamma_{\underline i}\gamma_{\underline r} \epsilon_L + \frac12 {\rm e}^{K/2} W \gamma_i \epsilon_R \label{delpsiiI} \\[2mm]
\delta \chi_L & = & \frac12 \gamma^r z^{i\prime}(r) \epsilon_R - \frac12 {\rm e}^{K/2} g^{i\bar \jmath} \overline{D_j W} \epsilon_L, \label{delchiI} 
\end{eqnarray}
where $ 
\partial_i \epsilon_L = 0$ was assumed.
Using also $\gamma_{\underline i} = \frac{1}{a} \gamma_i$, one can start solving the various equations.
From (\ref{delpsiiI}), splitting the complex superpotential into norm and phase 
\begin{equation}
W = |W| {\rm e}^{i \theta(r)}, \label{Wsplit} 
\end{equation}
we get the projector condition 
\begin{equation}
\gamma_{\underline r} \epsilon_L = \mp {\rm e}^{i \theta} \epsilon_R.
\label{proj0} 
\end{equation}
from which the BPS equation follows 
\begin{equation}
\frac{a^\prime}{a} = \pm |W|{\rm e}^{K/2} = \pm |{\cal Z}|.
\label{BPSI1} 
\end{equation}
The condition (\ref{proj0}) can be assembled into a real projector ${\cal P}$ (satisfying ${\cal P}^2 = {\cal P}$) acting on the full spinor as ${\cal P}\epsilon = \epsilon$: 
\begin{equation}
{\cal P} = \pm \frac{1}{2 \cos \theta} \left[\gamma_{\underline r} \pm \left(\cos \theta + i \gamma_5 \sin \theta\right)\right].
\label{projectorI} 
\end{equation}

The radial equation on the gravitino gives a differential equation on the spinor parameter 
\begin{equation}
\epsilon^\prime_L = -\frac{i}{2}A_r^B \epsilon_L \pm \frac12 {\rm e}^{K/2} |W| \epsilon_L.
\label{diffeps} 
\end{equation}
Here, the projector condition was used.
Moreover, one can use the projector condition inside (\ref{diffeps}), or take the charge conjugate (using the Majorana condition).
The two resulting equations are 
\begin{eqnarray}
\epsilon_R^{\prime} & = & - i \theta^{\prime} \epsilon_R -\frac{i}{2}A_r^B \epsilon_R \pm \frac12 {\rm e}^{K/2} |W| \epsilon_R, \label{projprime} \\[2mm]
\epsilon_R^{\prime} & = & \frac{i}{2}A_r^B \epsilon_R \pm \frac12 {\rm e}^{K/2} |W| \epsilon_R, \label{Majorana} 
\end{eqnarray}
whose consistency implies 
\begin{equation}
\theta^\prime = - A_r^{B} = \frac{i}{2}\left(z^{i\prime} 
\partial_i K -\bar z^{\bar \imath \prime} 
\partial_{\bar \imath} K \right).
\label{BPSextra} 
\end{equation}

The solution is given in terms of an arbitrary constant left-handed spinor parameter $\epsilon_0$ (hence the 1/2 supersymmetry preserved): 
\begin{equation}
\epsilon = a^{1/2} \left({\rm e}^{\frac{i}{2} \theta}\epsilon_0 \mp{\rm e}^{-\frac{i}{2} \theta}\epsilon_0^{C}\right).
\label{spinorI} 
\end{equation}
The BPS equation (\ref{BPSI1}) has been used to express the integration over the superpotential in terms of the warp factor and (\ref{BPSextra}) for the integration over $A_r^B$.
In the last formula we have used the definition of the charged conjugate spinor $\epsilon_{0}^C$ that follows from choosing the charge conjugation matrix $C = \gamma_{\underline r} \beta$, with $\beta = i \gamma_0$.
In this way the spinor $\epsilon$ satisfies the relation (\ref{proj0}).

Finally, using again the projector condition in the modulini equations, one finds the BPS equation for the scalar fields 
\begin{equation}
z^{i\prime} = \mp {\rm e}^{K/2+i\theta} g^{i\bar \jmath} \overline{D_{j}W}.
\label{BPSI2a} 
\end{equation}
We can now also remove the dependence on the phase $\theta$, by using the holomorphicity of the superpotential: 
\begin{equation}
\bar 
\partial W = 0 \qquad \Rightarrow\qquad \bar 
\partial |W| + i|W| \bar 
\partial \theta = 0.\label{holoW} 
\end{equation}
This implies that the BPS equations for the scalars can take the form: 
\begin{equation}
z^{i\prime} = \mp {\rm e}^{K/2} g^{i\bar \jmath} \left(2 
\partial_{\bar \jmath} |W| + 
\partial_{\bar \jmath} K |W|\right) = \mp 2g^{i\bar\jmath} 
\partial_{\bar \jmath}\left({\rm e}^{K/2}|W|\right) = \mp 2g^{i\bar\jmath} 
\partial_{\bar \jmath}|{\cal Z}|.
\label{BPSI2} 
\end{equation}
Summarizing, the supersymmetry conditions that have to be satisfied in order to specify the domain wall solutions are 
\begin{equation}
\fbox{$ 
\begin{array}{rcl}
    \displaystyle \frac{a^{\prime}}{a} & = & \pm |{\cal Z}|, \\[4mm]
    z^{i \prime} & = & \mp 2 g^{i\bar \jmath} 
    \partial_{\bar \jmath} |{\cal Z}|.
\end{array}
$} \label{summary} 
\end{equation}
The other consistency condition we derived from the solution of the  supersymmetry equations, namely (\ref{BPSextra}), is identically  satisfied.
This can be seen by using the fact that $\theta$ is the phase of the  superpotential, and by equating with (\ref{BPSextra}) the condition coming from its  definition,  $\theta^\prime = \frac{i}{2} \left(-\frac{W^{\prime}}{W}+ \frac{\overline W^{\prime}}{\overline W}\right)$.
The resulting equation is an identity upon using (\ref{summary}). 

Domain walls in $N=1$ supergravity were studied before in \cite{Cvetic:1996vr} in a more restrictive setting, so we compare our results.
First there is a change in the signature.
This implies that one should change the sign for every upper index and put an $i$ for each gamma matrix in the supersymmetry equations.
This can be re-absorbed in sending $W_{CV} \to -i W_{here}$.
Then there is a difference in the metric.
On the solution they have $A = B = a^{2}$, but also their radial coordinate $z$ is related to ours by $\frac{dr}{dz} = \sqrt{A}$.
Also, the decomposition of the superpotential in norm and phase implies that $\theta_{CV} = \theta$.
Altogether the equations in \cite{Cvetic:1996vr} become 
\begin{eqnarray}
\frac{a^{\prime}}{a} & = & \zeta {\rm e}^{K/2}|W|, \label{Cv1} \\[2mm]
T^{\prime} & = & -\zeta {\rm e}^{K/2+i \theta} g^{T\overline T} \overline{D_T W}, \label{Cv2} \\[2mm]
\theta^{\prime} & = & \frac{i}{2} \left(T^{\prime} 
\partial_{T} K - \overline T^{\prime} 
\partial_{\bar T} K\right).
\label{Cv3} 
\end{eqnarray}
It is easy to see that these equations are equivalent to the ones presented above, when restricted to a single scalar field flowing.

\subsection{A ``c-theorem" for N=1/2 domain walls}\label{sec:entropy}

We find it useful to present the domain wall equations in the form of renormalization group equations.
For this purpose we will describe the flow not as a function of $r$ coordinate but as a function of the warp factor.
We multiply Eq.~(\ref{BPSI2a}) on ${a\over a'}$ and find, using the Eq.~(\ref{BPSI1}) 
\begin{equation}
{a\over a'} z^{i\prime} = - g^{i\bar \jmath} {\overline{D_{j}W}\over \overline W}.
\label{renorm} 
\end{equation}
and 
\begin{equation}
\fbox{$ 
\begin{array}{rcl}
    \displaystyle a { 
    \partial z^i \over 
    \partial a} = - g^{i\bar \jmath} 
    \partial_{\bar \jmath} (K+\ln \overline W) 
\end{array}
$} \label{renorm1} 
\end{equation}
We will find below that in certain cases the domain wall BPS equations in the form (\ref{summary}) are useful.
Sometimes we will use equations in the form (\ref{renorm1}), particularly in the case of many moduli participating in the flow  and $W$ not crossing zero.

A careful inspection of the BPS conditions (\ref{summary}) shows that we can associate to the domain wall solutions a monotonic function, which therefore specifies the direction of the flow.
As it is known since the analysis of the 5-dimensional domain walls and the interpretation of such solutions as renormalization group flows in the context of the AdS/CFT correspondence, there is a c-function that is given by the first derivative of the logarithm of the warp factor $ c= ( \log a)^{\prime}$ \cite{Freedman:1999gp}.
By using the BPS equations we easily see that 
\begin{equation}
c^{\prime} = \left(\frac{a^{\prime}}{a}\right)^{\prime} = \pm |{\cal Z}|^{\prime} = \pm( z^{i\prime} 
\partial_{i}+ \bar z^{\bar \imath\prime} 
\partial_{\bar \imath})|{\cal Z}| = 4 g^{i\bar\jmath} 
\partial_{i}|{\cal Z}| 
\partial_{\bar \jmath} |{\cal Z}| \geq 0, \label{cfunc} 
\end{equation}
where the last inequality follows from the properties of the scalar metric.

There is, however, another interesting quantity that in most cases works as a c-function.
This is the square of the covariantly holomorphic superpotential, which is related to the square of the mass of the gravitino at a generic point of the moduli space 
\begin{equation}
|{\cal Z}|^2 = e^{K}|W|^2 = {1\over 3} M^{2}_{3/2}(z, \bar z).
\label{Mgravitino} 
\end{equation}
We can now look at the variations of $|{\cal Z}|^2$ along the flow.
In order to compute it, we actually evaluate $(|{\cal Z}|^2)^\prime = z^{i\prime} 
\partial_i |{\cal Z}|^2+ \bar z^{\bar \imath\prime} 
\partial_{\bar \imath}|{\cal Z}|^2 $ and use the BPS equation (\ref{BPSI2}).
After some easy algebra we get that 
\begin{equation}
(|{\cal Z}|^2)^{\prime} = \mp 8 \pi |{\cal Z}| g^{i\bar\jmath} 
\partial_i |{\cal Z}| 
\partial_{\bar \jmath}|{\cal Z}|.
\label{Sprime} 
\end{equation}
Due to the positive definiteness of the metric of the scalar fields, this expression is clearly either always increasing, or always decreasing along the flow, depending on the sign of the BPS equation, unless the flow crosses points where the superpotential vanishes.
For these special cases the BPS equations change sign as we cross the zeros of the superpotential and therefore $|{\cal Z}|^2$ is first decreasing until it reaches zero and then starts increasing again.

\subsection{BPS form of the action}\label{sub:BPSformS}

The supersymmetric domain walls studied above are static 1/2 BPS configurations of $N = 1$ supergravity.
In an analogous situation for static 1/2 BPS cosmic strings in N=1 supergravity \cite{Dvali:2003zh} it was possible to rewrite the gravitational action, including the boundary terms, in the BPS form.
The action therefore was given by some squares of the BPS conditions and total derivative terms.
A saturation of the BPS bound is equivalent to solving first order differential equations for unbroken supersymmetry.
The on shell action therefore is equal, up to a sign, to the total energy of the string.
Therefore the tension of the string, the energy per unit area, can easily be derived from the BPS action.
The metric of the cosmic string is $ds^2= -dt^2 + dz^2 + dr^2 + C^2(r) d\theta^2$ and the relation between the $-S$ on shell and a tension $\mu$ is 
\begin{equation}
-S^{\rm \, cosmic \, strings}_{\rm \, on \, shell} = \int dt \,dz \, \int dr d\theta \sqrt g T_0^0 = \int dt \,dz \, \mu ,
\end{equation}
where $\mu = 2 \pi \xi$ and $\xi$ is a FI term in the action.

For flat domain walls the metric is $ds^2= a^2(r)(-dt^2+dx^2+dy^2) + dr^2$.
Therefore one would expect the following relation between the action and the tension (the energy per unit of area) 
\begin{equation}
-S^{\rm \, domain\, walls}_{\rm \, on\, shell}= \int dt\, dx\, dy \, a^3(r) \, \sigma = \int d\tilde t\, d\tilde x\, d \tilde y \, \, \sigma . \label{expectation} 
\end{equation}
In the case of cosmic strings $t$ and $z$ have flat metric, therefore the volume of the area is just $\int dt\, dx$.
However, for our domain walls we have to distinguish the physical volume from coordinate volume and we have to ignore (divide by) the physical volume which is $\int dt\, dx\, dy \, a^3(r)= \int d\tilde t\, d\tilde x\, d \tilde y $.

In \cite{Cvetic:1996vr} the tension of the BPS domain walls was calculated using a rather complicated Nester construction.
The formula for the tension/energy density of the wall between two AdS critical points is given for two cases: in one case the superpotential does not vanish between critical points, in the other case it vanishes at some intermediate point:
\begin{itemize}
\item 2 AdS vacua, $W\neq 0$ in between 
\begin{equation}
    \sigma = {2\over \sqrt 3} \left(|\Lambda|_+^{1/2} - |\Lambda|_-^{1/2}\right) \label{1} ,
\end{equation}
\item 2 AdS vacua, $W= 0$ in between 
\begin{equation}
    \sigma = {2\over \sqrt 3} \left(|\Lambda|_+^{1/2} + |\Lambda|_-^{1/2}\right) \label{2} .
\end{equation}
\end{itemize}
Here we show that for domain walls we can put the supergravity action in the standard BPS form, namely as the integral of the square of the BPS conditions plus boundary terms.
We start with the gravitational action supplemented by the Gaussian curvature at the boundaries.
\begin{eqnarray}
S &&= \int d^{4}x \,\sqrt{-g}\left(\frac12 R - g_{i \bar \jmath} 
\partial_{\mu}z^i 
\partial^\mu \bar z^{\bar \jmath} - V(z, \bar z)\right)\nonumber\\
&&+ \int d^3 x \Big(\sqrt {-\det h}\; K| _{r=+\infty} - \sqrt {-\det h}\; K| _{-\infty}\Big) \, .
\label{action1} 
\end{eqnarray}
Here $h$ is the metric at the boundaries.
In our case the boundaries are at $r=const$ surface.
Therefore $\sqrt {-\det h}=a^3$ and $K= g^{ab} K_{ab} $ where $g^{xx}=g^{yy} = g^{tt}= a^{-2}$ and $K_{ab} = {1\over 2} n^\mu { 
\partial g_{ab}\over 
\partial x^\mu}$ and $n^\mu = {1\over \sqrt {g_{rr}}} \Big ({ 
\partial \over 
\partial r}\Big ) ^\mu $.
One starts by plugging (\ref{scalarI}) in the Einstein term and using that all the quantities depend only on the radial coordinate.
Then, we try to re-combine the various terms in the form of the BPS equations (\ref{BPSI1})--(\ref{BPSI2a}) plus some boundary term.
The action (\ref{action1}) becomes 
\begin{eqnarray}
S &&= \int d^3x dr \left[- a^3 \left(3 \left(\frac{a^\prime}{a}\right)^2 + 3 \frac{a^{\prime\prime}}{a}\right)-a^3 g_{i\bar\jmath}z^{i\prime}\; \bar z^{\bar \jmath \prime}+ 3 a^3 | {\rm e}^{K/2} W|^2- a^3 {\rm e}^K g^{i\bar \jmath} D_i W \overline{D_j W}\right]\nonumber \\
&&+ \int d^3 x \, ( a^3)'| _{r=+\infty} - (a^3)'| _{r=-\infty}) \ .
\end{eqnarray}
This can now be recast in the form of BPS equations squared plus additional boundary terms as 
\begin{eqnarray}
S &=& \int d^3 x \, dr \left[3 a^3 \left(\frac{a^\prime}{a}\pm {\rm e}^{K/2}|W|\right)^2 - a^3 g_{i\bar\jmath}\left(z^{i\prime}\mp e^{K/2+i \theta}g^{i \bar k}\overline{D_k W}\right) \left(\bar z^{j\prime}\mp e^{K/2-i \theta}g^{\bar \jmath k}D_k W\right)\right] \nonumber \\[2mm]
&\mp& \int d^3 x \, dr \left[2(a^3)^\prime |{\rm e}^{K/2}W| + a^3 z^{i\prime} 
\partial_i (e^{K/2}W){\rm e}^{-i\theta}+a^3 \bar z^{\bar \imath}\bar 
\partial_i (e^{K/2}\bar W){\rm e}^{i\theta}+ a^3 (e^{K/2})^{\prime} |W|\right] \label{BPSform0} 
\end{eqnarray}
Finally, the second line becomes a total derivative by using (\ref{holoW}), and the total action reads 
\begin{eqnarray}
S &=& \int d^3 x \, dr \left[3 a^3 \left(\frac{a^\prime}{a}\pm |{\cal Z}|\right)^2 - a^3 g_{i\bar\jmath}\left(z^{i\prime}\mp 2g^{i\bar k} 
\partial_{\bar k}|{\cal Z}|\right)\left(\bar z^{j\prime}\mp 2g^{\bar\jmath k} 
\partial_k|{\cal Z}|\right)\right] \nonumber \\[2mm]
&\mp& \int d^3 x \, dr \frac{d}{dr}\left[a^3\left(2 |{\cal Z}|\right)\right], \label{BPSformI} 
\end{eqnarray}
where we recovered the BPS equations in the first line and we have a boundary term contribution in the second line.
One can define the energy density functional $\sigma$, since we have a static solution, so that $S|_{on shell}= - \int d^3 x \, a^3 \, \sigma$.
If the superpotential does not vanish between the critical points we find 
\begin{eqnarray}
\int d^3 x \, a^3 \, \sigma &&= \int d^3 x \, dr \left[- 3 a^3 \left(\frac{a^\prime}{a}+ |{\cal Z}|\right)^2 + a^3 \Big |z^{i\prime}- 2g^{i\bar k} 
\partial_{\bar k}|{\cal Z}| \Big |^2 \right ]\nonumber\\
&& + \int d^3 x ( a^3 2 |{\cal Z}|_ {r=+\infty}- a^3 2 |{\cal Z}|_ {r=-\infty}). \label{BPSform2} 
\end{eqnarray}
In case that $|{\cal Z}|_ {r=+\infty}- |{\cal Z}|_ {r=-\infty}>0$ we have to make a choice of the sign in BPS equations such that the tension is positive, as shown in Eq.~(\ref{BPSform2}).
Thus when $|{\cal Z}|$ does not vanish between fixed points and the BPS equation of motion are satisfied with the upper sign the energy density of the wall is given by 
\begin{eqnarray}
\sigma_{{}_{{\cal Z}\neq 0}} = 2 ( |{\cal Z}|_ {r=+\infty}- |{\cal Z}|_ {r=-\infty}). \label{tension1} 
\end{eqnarray}
Note that $\Lambda = 3 |{\cal Z}|^2$.
So our answer is exactly as in eqs.~(\ref{1}), in agreement with \cite{Cvetic:1996vr}.
It was explained in \cite{Cvetic:1996vr} that the BPS domain walls with this tension are related to CDL bubbles with infinite radius.

Now we assume that there is a solution of the BPS equations where the superpotential vanishes at some point $r_0$ between the AdS critical points.
The boundary integral in Eq.~(\ref{BPSformI}) must be split into two terms since one part of the flow proceeds with one sign in the BPS equations and for the other part, after ${\cal Z}=0$ point, the sign is flipped.
Therefore the answer for the energy density in this case is 
\begin{eqnarray}
\sigma_{{}_{{\cal Z}(r_0)=0}} = 2 \Big ( -(|{\cal Z}|_ {r_0}- |{\cal Z}|_ {r=-\infty}) + |{\cal Z}|_ {r=+\infty}-|{\cal Z}|_ {r_0} \Big )= 2 ( |{\cal Z}|_ {r=+\infty} + |{\cal Z}|_ {r=-\infty}). \label{tension2} 
\end{eqnarray}
Noting that $\Lambda = 3 |{\cal Z}|^2$, we have derived the correct tension as in Eq.~(\ref{2}) in agreement with \cite{Cvetic:1996vr}.
This case, according to \cite{Cvetic:1996vr} is not related to CDL tunneling.

We would like to conclude this subsection by the observation that our new derivation of the BPS domain wall tension follows the pattern discovered for the BPS cosmic string in \cite{Dvali:2003zh}.
We find the derivation using the BPS form of the action significantly more elegant than using the Nester construction.

\section{BPS domain walls between vacua not separated by barriers} 

In this section we are going to give some examples of supersymmetric domain wall solutions for some of the generic models which capture the most important features of the landscape of flux vacua.
We start with the simple scenario with just one K\"ahler modulus and all the complex structure moduli frozen: the KKLT model \cite{Kachru:2003aw}.
This model has one supersymmetric AdS vacuum at a finite point of the volume modulus and a Minkowski vacuum at infinite value of the volume modulus.
We will solve the BPS equations numerically and display the corresponding domain wall.
After this simple example we show a more involved setup where also the complex structure moduli and the axion-dilaton are allowed to flow: the $STU$ models.
These include a wide variety of models (and also the KKLT model) as a special case.
The case which we will solve numerically has two different AdS critical points at finite distance in the moduli space and three scalars are flowing between the critical points, two of them proportional to each other.
Finally we focus on a model where the effect of the K\"ahler deformations is neglected, but the form of the complex-structure deformations is more involved.
The flux vacua in such models are described by a simple set of ``new attractor equations'' \cite{Kallosh:2005ax} relating the points in the moduli space where the moduli are stabilized by fluxes\footnote{These new attractor equations for IIB vacua have also been extended to the case of non-K\"ahler compactifications of the heterotic theory \cite{DallAgata:2006nr}. 
Similar extensions for the black hole case have been discussed in \cite{Hsu:2006vw}.}.
These flux vacua attractors have some resemblance as well as some differences with the black hole attractor equations \cite{FKS}.
We will focus here on a model of a Calabi--Yau threefold obtained as a hypersurface in $WP^{4}_{1,1,1,1,2}$.
This model has two different AdS critical points \cite{Giryavets:2005nf} at finite distance in the moduli space.
We will find out how four scalars in this model are flowing from the Landau--Ginsburg AdS critical point towards the near conifold region of the moduli space.

\subsection{The simplest KKLT model}\label{sec:KKLTmodel}

The KKLT model \cite{Kachru:2003aw} suggests a way to generate meta-stable de Sitter vacua in string theory starting from supersymmetric Anti de Sitter minima with all moduli stabilized due to fluxes and using non-perturbative effects, such as gaugino condensation and/or D3-instantons.
The uplifting of these stabilized AdS vacua to dS vacua is possible under certain conditions by the addition of positive energy density sources, like anti-D3 branes.
The simplest realization has a single volume modulus $\rho={\rm e}^{\frac{2}{\sqrt3}\phi}+ i \alpha$.
After all the complex structure moduli have been fixed by the fluxes, the superpotential for the volume modulus (whose K\"ahler potential is $K = -3 \log(\rho+\bar\rho)$) is 
\begin{equation}
W = W_0+ A {\rm e}^{-a \rho}, 
\end{equation}
with $W_0$, $A$ and $a$ parameters depending on the details of the setup.

\begin{figure}[hbt]
\label{KKLTpot} \centering 
\includegraphics[scale=0.9]{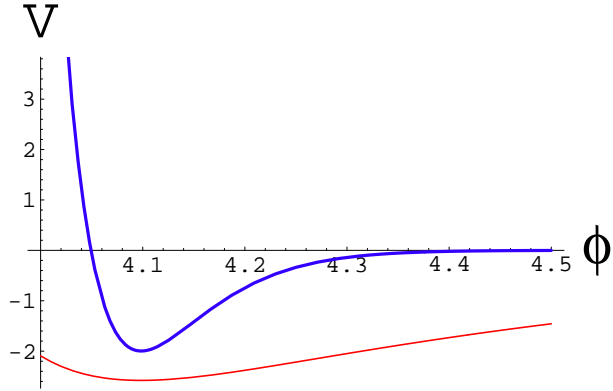} \caption{\small KKLT potential $V$ (blue upper curve) and ${\cal Z}$ (red curve below) as functions of the modulus.
One can see that the potential has an AdS minimum at $\phi = \phi^{*}\simeq 4.09854$ and tends to Minkowski limit at large $\phi$.
${\cal Z}$ is always negative in the interval between AdS and Minkowski vacua.
Potential energy density $V$ is shown in units $10^{{-15}} M_{p}^{4}$, whereas the covariantly holomorphic superpotential ${\cal Z}$ is shown in units $10^{{-8}} M_{p}^{3}$.}\label{plot1} 
\end{figure}

The corresponding potential, before uplifting, has one AdS supersymmetric critical point at finite volume and an asymptotic Minkowski vacuum at infinity of the volume.
It is easy to check that we can consistently set the axion $\alpha$ to zero, since this is a fixed point for any value of the volume $\phi$.
For the sake of simplicity we will also take the parameters $A$, $a$ and $W_0$ to be real (and positive the first two $A = 1$, $a = 0.1$, negative the last one $W_0 = -0.0001$).
The potential for this case and the covariantly holomorphic superpotential ${\cal Z}$ are given by Fig.~\ref{KKLTpot} as functions of the volume modulus.

One finds that for these parameters during the total flow the covariantly holomorphic superpotential ${\cal Z}$ is negative and approaches zero at infinity.
We can reduce the problem to a one-dimensional gradient-flow, where the BPS equations follow by using 
\begin{equation}
{\cal Z}|_{\alpha = 0} = -|{\cal Z}| = \frac{{\rm e}^{-\sqrt3 \phi}}{\sqrt{8}}\left (W_0 + A\,{\rm e}^{-a {\rm e}^{\frac{2}{\sqrt3}\phi}}\right ) \ .
\end{equation}
We solve numerically the flow equations in the opposite direction (for technical reasons), starting with the AdS minimum towards the infinitely far Minkowski.
For this purpose we have to use (\ref{summary}): 
\begin{eqnarray}
a^\prime(r)&=&- a(r) |{\cal Z}(r)|,\\[4mm]
\phi^\prime (r) &=& 
\partial_\phi |{\cal Z}(r)|.
\end{eqnarray}
The AdS critical point is at $\phi = \phi^{*}\simeq 4.09854$, and for values slightly above $\phi^{*}$ we get a domain wall solution that drives $\phi \to \infty$ and ${\cal Z} \to 0$.
In this case we obviously have just one basin of attraction and the attractor point is the supersymmetric AdS vacuum.
Asymptotically, at large $r$, the value of $\phi$ grows, and $|{\cal Z}|$ becomes 
\begin{equation}
|{\cal Z}| \sim c \,{\rm e}^{-\sqrt3 \phi}, \qquad c = |W_0|/\sqrt8>0.
\label{Zas} 
\end{equation}
We can therefore solve the BPS equations above in this limit by 
\begin{equation}
a \sim a_0 (r - r_0)^{\frac13}, \qquad \phi \sim \frac{1}{\sqrt3} \log (3c (r-r_{0})).
\label{aasint} 
\end{equation}
It should be noted that 
\begin{equation}
\frac{a^{\prime}}{a} \sim \frac{1}{3(r-r_0)}, \label{mink} 
\end{equation}
which means that the curvature of the space-time (\ref{scalarI}) vanishes asymptotically, as it should be in Minkowski space.
\begin{figure}[hbt]
\centering 
\includegraphics[scale=0.75]{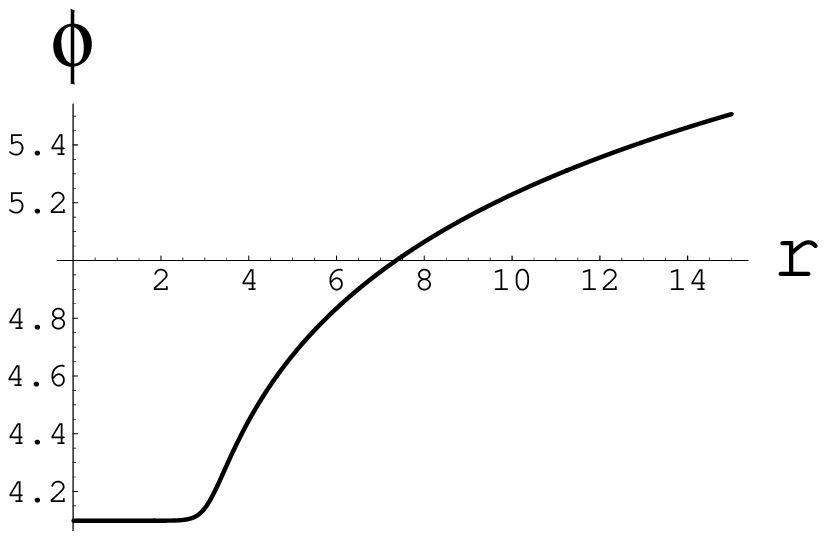}\hskip 0.5cm 
\includegraphics[scale=0.75]{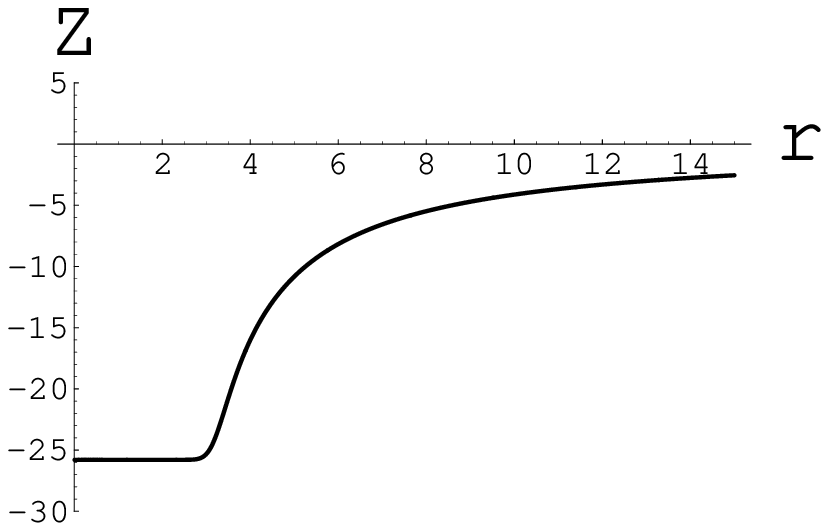} \vskip 0.5cm 
\includegraphics[scale=0.75]{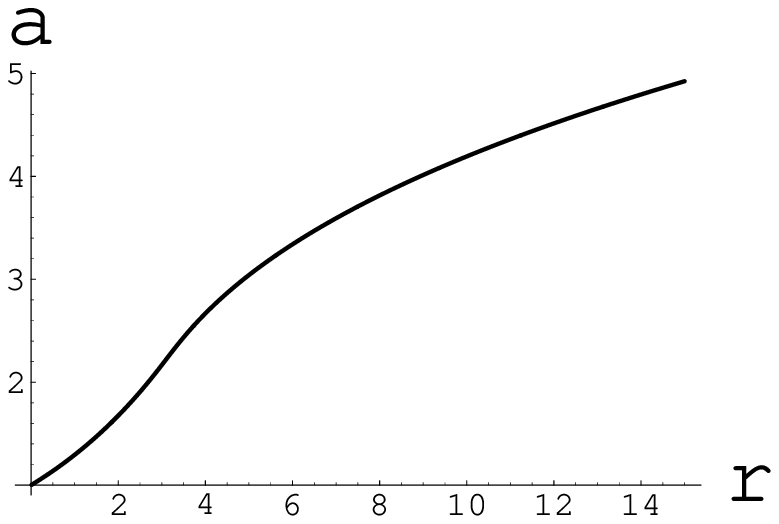}\hskip 0.5cm 
\includegraphics[scale=0.75]{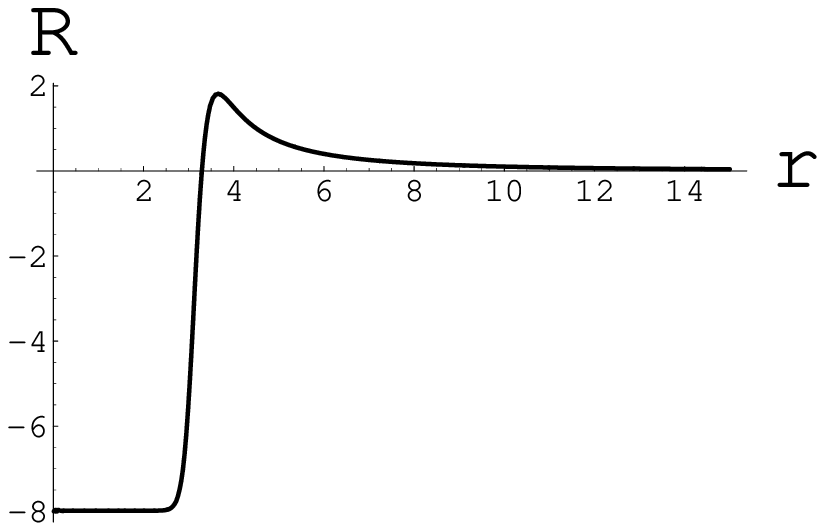} \caption{\small The KKLT wall interpolating between the AdS and flat Minkowski space which are at infinite distance from each other in the moduli space.
We plot the scalar $\phi$, the covariantly holomorphic superpotential ${\cal Z}$ (in units $10^{{-9}}$), the warp factor $a$ and the curvature $R$ (in units $10^{{-15}}$), all as functions of $r$.
Here $r$ is the coordinate of the domain wall configuration given in Eq.~(\ref{typeI}); we show $r$ in units $10^{7}$.
} \label{KKLTabcd}
\end{figure}

In Fig.~\ref{KKLTabcd} we plot the wall: first on the top right we have a scalar interpolating between the AdS value $\phi = \phi^{*}\simeq 4.09854$ and infinity.
On the top right we plot ${\cal Z}$ as the function of wall coordinate $r$.
It is always negative but here we see clearly that it takes a constant negative value at AdS and starts interpolating towards zero at Minkowski vacuum.
The next plot on the bottom left shows the warp factor $a(r)$.
Finally on the right bottom we show the curvature of the domain wall: it is vanishing at large values of modulus in the flat space and grows towards the decreasing values of modulus and sharply drops down to its negative AdS value upon reaching the AdS vacuum.

From this example we learn that in the absence of a barrier between the AdS and Minkowski vacua there exists a static domain wall where the flow of the volume modulus from the infinite value at Minkowski towards the finite value at AdS is controlled by 1/2 of unbroken supersymmetry.
The tension of this wall is equal to $\sigma= {2\over \sqrt 3} |\Lambda|^{1/2}$ where $\Lambda$ is the cosmological constant of the AdS minimum.

\subsection{$STU$ models}\label{sec:T6Z2Z2}

The previous example had only one AdS vacuum.
However, in the landscape of flux compactifications there are also examples with multiple AdS vacua and therefore it is interesting to display an example with multiple basins of attraction to different AdS vacua and the supersymmetric domain wall solutions interpolating between them.
In order to find a workable example, we need an analytic superpotential admitting multiple AdS vacua.
Such an example is provided by various models of toroidal orbifolds analyzed in \cite{Lust:2005dy}.
The generic moduli content is given by the axio/dilaton $S$ and by the complex-structure moduli $U^{i}$ and 3 K\"ahler moduli $T^i$, with K\"ahler potential 
\begin{equation}
K = - \log (S+\bar S) - \sum_{i=1}^{3} \log (T^i + \overline T^i) - \sum_{i = 1}^{\stackrel{h_{(2,1)}}{\rm untwisted}} \log (U^i + \overline U^i).
\label{KahlerT6} 
\end{equation}
For the sake of simplicity we focus on the case of one untwisted complex structure modulus, as for instance in the ${\mathbb Z}_2 \times {\mathbb Z}_3$, ${\mathbb Z}_2 \times {\mathbb Z}_{6}$ orbifolds or in the ${\mathbb Z}_{6 - II}$ orbifold with $SU(2)^{2} \times SU(3) \times G_2$, $SU(3) \times SO(8) $, $SU(2)^{2} \times SU(3) \times SU(3)_{\natural}$ or $SU(2) \times SU(6)$ lattice.
The superpotential is a polynomial in the complex-structure and dilaton moduli that comes from the fluxes, added by some non-perturbative exponential terms involving the K\"ahler-structure moduli.
If we consider the cases with just one complex structure modulus, the full untwisted superpotential reads \cite{Lust:2005dy}
\begin{equation}
W = \displaystyle a + b \,U + c\, S + d\, SU+ \sum_{i=1}^{3} g_i\, {\rm e}^{-h_i T^i}.
\label{supoT6} 
\end{equation}
Here the parameters $a,b,c$ and $d$ are related to the fluxes (they are the coefficients of the real sections of the 3-cycles), while the parameters $g_i$ and $h_i$ are to be associated with non-perturbative effects.
This potential includes the one-modulus KKLT scenario as a special example.

We can then look for supersymmetric vacua, satisfying $D_S W = 0$, $D_U W = 0$ and $D_{T^i} W = 0$.
For simplicity, let us consider the case that $g_i = g$, $h_i = h$ and look for vacua where $T^i = T$.
Then, the supersymmetry conditions become 
\begin{eqnarray}
c+ d\, U & = & \frac{1}{S+\bar S} W, \label{DSW} \\[2mm]
b+ d\, S & = & \frac{1}{U+\bar U} W, \label{DUW} \\[2mm]
- gh\, {\rm e}^{-h T} & = & \frac{1}{T+\bar T} W.
\label{DTW} 
\end{eqnarray}
A generic solution for $W \neq 0$ of these equations requires that 
\begin{equation}
U = \frac{c}{b}\, S = k\,S, \label{relUS} 
\end{equation}
and if we further look for the case where Im$T$ = Im$S$ = 0, then we also obtain that 
\begin{eqnarray}
{\rm Re}\,T & = & -\frac{1}{h} \log \left[-\frac{a-kd \, {\rm Re}\,S^2}{3g}\right], \label{cond1} \\[2mm]
{\rm Re}\,T & = & \frac{3c{\rm Re}\,S(b + d\, {\rm Re}\,S)}{h(a b- cd \,{\rm Re}\,S^2)}.
\label{label3} 
\end{eqnarray}
When the values of the fluxes are chosen appropriately, one can obtain 2 distinct vacua at values of  $s = {\rm Re} S > 0$ and $t = {\rm Re} T > 0$.
One instance is the choice $a = 1, b = 2, c = 4, d = -1, g = -10, h = 1/100$.
The critical points are at $\{s = 0.496855, t= 299.992\}$ and $\{s=0.749598,t=264.799\}$.
(The value of the size of the manifold is tuned by $1/h$.
The smaller $h$, the bigger $t$.
Increasing $g$ on the other hand allows for smaller values of $s$.)

We can then use the BPS equations to obtain the interpolating solution between them.
Having seen how the critical points are obtained, we would like to reduce the problem to a two-dimensional system depending only on $s = {\rm Re}S$, $t = {\rm Re}T$.
Both critical points lie on the section of the moduli space where all the axions are set to zero.
This condition can be kept consistently all along the flow.
When the axions are vanishing, $W = |W|$ and therefore $\theta = 0$.
This condition can be preserved along the flow as $\theta^{\prime} = 0$ if the axions are vanishing.
This also implies, since the metric is diagonal, that the real and imaginary parts decouple consistently.
Given the symmetry of the potential, we can also assume that $T^i = T$ all along the flow.
The BPS equations therefore reduce to 
\begin{eqnarray}
s^{\prime} & = & \mp 8s^2 
\partial_s |{\cal Z}|(s,t,u) , \nonumber \\[2mm]
t^{\prime} & = & \mp 8t^2 
\partial_t |{\cal Z}|(s,t,u) , \label{boh} \\[2mm]
u^{\prime} & = & \mp 8u^2 
\partial_u |{\cal Z}|(s,t,u) , \nonumber 
\end{eqnarray}
and, by using (\ref{relUS}), one can prove that the differential equations for $u$ and $s$ are proportional to each other.
This means that we can further reduce the problem to the description of the gradient flows for the flow potential $|{\cal Z}|(\phi_i)$, where $s = {\rm e}^{\sqrt8 \phi_1}$ and $t = {\rm e}^{\sqrt8 \phi_2}$.
In these variables, the BPS equations become 
\begin{equation}
\left\{ 
\begin{array}{rcl}
    \phi_i^{\prime} & = & \mp 
    \partial_{\phi_i} |{\cal Z}|(\phi_i) \\[2mm]
    \displaystyle \frac{a^{\prime}}{a} & = & \pm |{\cal Z}|(\phi_i) 
\end{array}
\right.\label{BPSeff} 
\end{equation}
where the absolute value of the central charge reads 
\begin{equation}
|{\cal Z}|(\phi_i) = \sqrt{\frac{b}{32 c}} {\rm e}^{\sqrt8(-\phi_1-3\phi_2)}\left(a + \frac{c}{b}{\rm e}^{\sqrt8\phi_1}(2b + d e^{\sqrt8 \phi_1})+ 3 g {\rm e}^{h {\rm e}^{\sqrt8\phi_2}}\right).
\label{Z} 
\end{equation}
Notice the double exponentials due to the non-perturbative terms in the K\"ahler moduli.

Given this flow potential, we can look for the special case where we have two distinct critical points, for instance when the parameters are chosen as above: $a = 1, b = 2, c = 4, d = -1, g = -10, h = 1/100$.
For these parameter values (\ref{Z}) shows two supersymmetric AdS critical points with different values of the cosmological constant and an asymptotically Minkowski vacuum.

\begin{figure}[h!] \centering 
\includegraphics[scale=0.9]{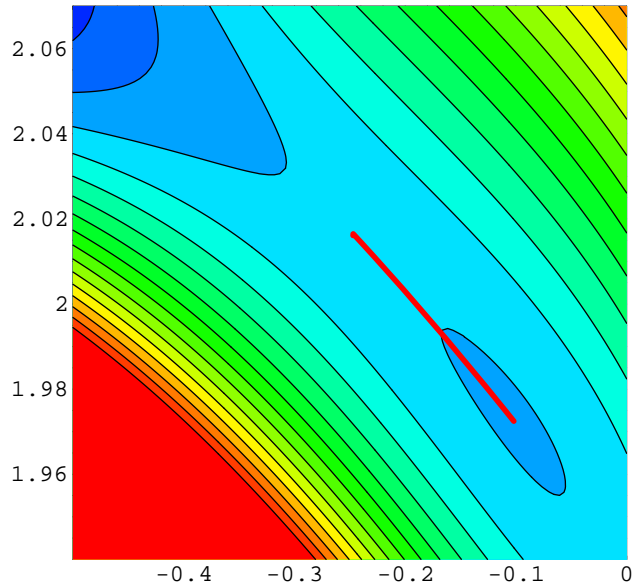} \caption{\small Here is the contour plot of the covariantly holomorphic superpotential ${\cal Z}(\phi_1,\phi_2)$ and the gradient flow between the two critical points.
${\cal Z}(\phi_1,\phi_2)$ is everywhere positive between AdS critical points, it vanishes only at the asymptotically Minkowski vacuum.} \label{fig:contplotW} 
\end{figure}

\begin{figure}[h!]
\centerline{ 
\includegraphics[scale=.7]{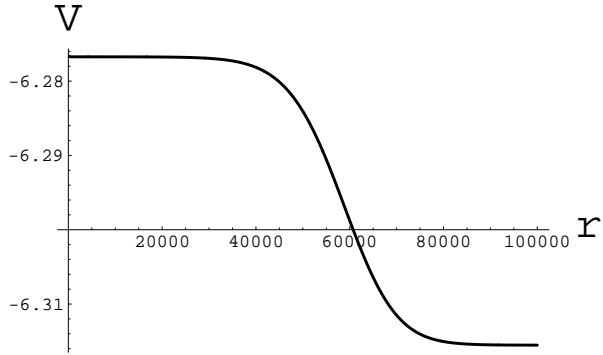} } \caption{\small Flow of the potential $V(\phi_{1}(r),\phi_{2}(r))$.}\label{fig:STUpot} 
\end{figure}

Although the BPS equations can be written in a closed form, the integration can only be done numerically.
In Fig.~\ref{fig:contplotW} we show the contour plot of ${\cal Z}$, where the $\phi_1$ modulus (proportional to the complex structure and to the dilaton $s = {\rm e}^{\sqrt8 \phi_1}$) is on the horizontal axis and $\phi_2$, proportional to the K\"ahler modulus $t = {\rm e}^{\sqrt8 \phi_2}$ is on the vertical one.
We plot the total potential on Fig.~\ref{fig:STUpot}.
We can clearly see from the pictures that there are two distinct critical points, and we have plotted the numerical solution interpolating between them.
As it is clear from the picture, the two critical points have different behavior.
Using the nomenclature coming from 5d, one is approached by the flow in an ultra-violet direction (i.e.~the warp factor goes as $a \sim {\rm e}^{r} \to +\infty$, for $r \to \infty$), the other in an infra-red (i.e.~the warp factor goes as $a \sim {\rm e}^{r} \sim 1$ for $r \to - \infty$).
An ultra-violet critical point is a minimum of the absolute value of the superpotential (minimum with positive $W$ or maximum with negative $W$), an infra-red critical point is a maximum.
This implies that there is no need for the superpotential to cross zero.
The warp factor indeed shows no critical points, as it always increases along the flow.
The superpotential will necessarily cross zero along the flow and therefore the BPS equations will change sign only when the interpolation must be done between two infra-red (two ultra-violet critical points are excluded as the wall should be singular to interpolate between them).
This is the only way to obtain a smooth domain wall interpolating between them.
A similar phenomenon occurred in the realization of the Randall--Sundrum scenario with a thick brane in \cite{Behrndt:2001km}.
We will have examples of such domain walls in Sec.~\ref{sec:subcritical}.

From Fig.~\ref{fig:contplotW} we can also clearly see the two basins of attraction for the AdS and Dine-Seiberg Minkowski critical points and the AdS repeller (with a marginal line of attraction separating the two basins).

This shows that there is one main basin of attraction for the AdS vacuum with the larger absolute value of the cosmological constant.
There is a line of attraction for the other critical point, which is really a saddle point of the flow potential.
Finally there is another basin of attraction for the values of the moduli that leads to the decompactification of the space.
The size modulus $t$ runs to infinity at the same time as the dilaton goes to zero, so we get a decompactification of the space in the weak coupling regime.
This example shows that the landscape attractors are different from the black-hole attractors.
For the latter, supersymmetry vacua are always minima of the superpotential $W$ and therefore one always has clear distinct basins of attractions and there are no repeller-like vacua. Moreover, if one is looking at the 5-dimensional case there is always only one basin of attraction for the vector-multiplet moduli.

\subsection{Example of a Calabi--Yau hypersurface}

The panorama of the landscape can be further refined by considering a representative example in the class of type II compactifications on Calabi--Yau threefolds, defined as hypersurfaces in a weighted projective space.
When fluxes are active for the RR and NS three-form field strengths on the orientifold of these hypersurfaces, a non-trivial superpotential is induced for the dilaton and complex structure moduli, leading to the many interesting physical properties that have driven a massive research activity (for a review and bibliography see \cite{Grana:2005jc}).

These models are particularly relevant even for a single deformation parameter because of the results obtained in \cite{Giryavets:2004zr} showing that, quite unexpectedly, the density of flux vacua is not uniform in moduli space, but rather it peaks around the conifold singularity, $\psi=1$.
Indeed, upon extensive numerical computations, it has been shown that the conifold region plays the role of ``attractor" basin and it serves as accumulation region for flux vacua.
This supports the analytic predictions made in \cite{Ashok:2003gk}.
On the other hand, also the Landau--Ginzburg singularity at $\psi=0$ enjoys very peculiar properties, deriving from it being a fixed point of the modular group.
In general, it is the special point in moduli space where the number of vacua with vanishing superpotential $W=0$ and discrete symmetries can be of the same order as the total number of vacua (when non-zero) \cite{DeWolfe:2004ns}.
For our purposes, it will be convenient for computations because of our need to use perturbative expansions for the CY periods, that are known around this point \cite{Klemm:1992tx}.
Thus, in spite of neglecting the K\"ahler structure moduli that in principle would be present in these systems, this case is particularly interesting as a representative of the large amount of possible flux vacua that are making the landscape so rich.

Another important related feature is that one may encounter multiple attraction basins \cite{Giryavets:2005nf}.
These consist of distinct supersymmetric vacua that can be found within a finite region in moduli space, or ``area code'', for fixed values of the $F_{(3)}$ and $H_{(3)}$ fluxes.
We thus proceed to identify such a situation and to construct the domain wall solution interpolating between different area codes.
As a start, we quickly recall some necessary formulas, while referring for instance to \cite{DeWolfe:2004ns} for the relevant lore and more conventions.

One examines in general the $h_{(2,1)}$ complex structure deformations.
As usual, one chooses a symplectic basis $\{A^a,B_b\}$ for the $b_3=2 h_{(2,1)}+2$ homology cycles ($a,b=1,\ldots ,h_{(2,1)}+1$), with dual cohomology elements $\{\alpha_a,\beta^b\}$ such that 
\begin{equation}
\int_{A^a} \alpha_b = \delta^a_b\, ,\qquad \int_{B_b} \beta^a =- \delta^a_b\, ,\qquad \int \alpha_a \wedge \beta^b =\delta^b_a\, , 
\end{equation}
and one assembles into a symplectic vector the periods of the unique holomorphic three-form $\Omega$, 
\begin{equation}
X^a=\int_{A^a}\Omega ,\qquad {\cal F}_b=\int_{B_b}\Omega ,\qquad \Pi(X) = ({\cal F}_a,X^a)\, .
\end{equation}
The $X^a$ are projective coordinates on the complex structure moduli space and ${\cal F}_b= 
\partial_b {\cal F} (X)$, where ${\cal F}(X)$ is the holomorphic prepotential.

The K\"ahler potential for the complex-structure and axion-dilaton moduli space is given by 
\begin{equation}
K=-\log\left(i\int_M \Omega\wedge\bar\Omega \right)-\log(-i(\tau-\bar\tau))=-\log\left(-i\Pi^\dagger\cdot\Sigma\cdot\Pi \right)-\log\left(-i(\tau-\bar\tau)\right)\, , 
\end{equation}
where $\tau$ is the axion-dilaton and $\Sigma=\left( 
\begin{array}{cc}
0 &1\\-1&0 
\end{array}
\right)$.
The superpotential induced by non-trivial 3-form fluxes reads 
\begin{equation}
W=\int_M G_{(3)}(\tau)\wedge\Omega(X)=(2\pi)^2\alpha^\prime (f-\tau h)\Pi(X)\, , 
\end{equation}
where $G_{(3)}=F_{(3)}-\tau H_{(3)}$ and $(f,h)$ are the integer valued flux components 
\begin{equation}
F_{(3)}=-(2\pi)^2\alpha^\prime(f_a\alpha^a+f_{a+h_{2,1}+1}\beta_a)\, ,\quad H_{(3)}=-(2\pi)^2\alpha^\prime(h_a\alpha^a+h_{a+h_{2,1}+1}\beta_a)\, .
\end{equation}
We will take for convenience $(2\pi)^2\alpha^\prime=1$.

Let us consider in the following $M$ to be the Calabi--Yau locus defined by the equation 
\begin{equation}
\sum_{i = 1}^{4}x_{i}^{6}+ 2x_0^3 - 6\psi\,x_0 x_1 x_2 x_3 x_4 = 0, \qquad x_i \in WP^{4}_{1,1,1,1,2} .
\end{equation}\
The supersymmetric vacua are found by solving the ``F--flatness" condition, that is a (K\"ahler covariantized) extremization with respect to $\tau$ and the single complex-structure modulus $\psi$, 
\begin{equation}
D_\tau W=( 
\partial_\tau W + 
\partial_\tau K W)=0\, ,\qquad D_a W=( 
\partial_a W + 
\partial_a K W)=0\, , 
\end{equation}
leading to the equivalent conditions 
\begin{equation}
(f-\bar\tau h)\cdot\Pi(X)=(f-\tau h)\cdot( 
\partial_\psi \Pi+ 
\partial_\psi K\,\Pi)=0\, .
\end{equation}
Alternatively, one can solve the flux vacua attractor equations \cite{Kallosh:2005ax}.
One finds that the moduli space of this model presents Landau--Ginzburg, conifold and large complex structure singular points respectively at $\psi=0$, $\psi=1$ and $\psi\to\infty$ \cite{Giryavets:2005nf}.

The monodromy group, $\Gamma$, of the complex structure moduli space has two generators: $A$, which generates phase rotations $\psi \to \alpha \psi$ with $\alpha = \exp (2 \pi i /6)$ around the Landau--Ginzburg point at $\psi = 0$, and $T$ which corresponds to the logarithmic monodromy ${\cal F}_2 \to {\cal F}_2 + X^2$ around the conifold singularity $\psi = 1$.
By itself, $A$ generates a ${\bf Z}_6 \subset \Gamma$ subgroup, with an associated fixed point at $\psi = 0$; $T$, on the other hand, is of infinite order.
The monodromy matrix $A$ generates rotations by a root of unity around $\psi =0$: 
\begin{equation}
A \Pi(\psi) = \alpha \Pi(\alpha\psi) 
\end{equation}
and is explicitly given by 
\begin{equation}
A=\left( 
\begin{array}{cccc}
    1 & -1 & 0 & 1\\
    0 & 1 & 0 & -1\\
    -3 & -3 & 1 & 3\\
    -6 & 4 & 1 & -3\\
\end{array}
\right) .
\end{equation}
The periods admit, for $|\psi| < 1$, an expansion in a Picard--Fuchs basis 
\begin{equation}
w_{i}(\psi) = {(2\pi i)^{3}\over 6}\sum_{n=1}^{\infty}{\exp({5\pi i\over 6}n)\Gamma({n\over 6}) \over \Gamma(n)\Gamma(1-{n\over 6})^3\Gamma(1-{n\over 3})} \left({6\alpha^{i}\over 2^{1/3}}\right)^{n}\psi^{n-1}\, , 
\end{equation}
where $\alpha$ is the $6^{\rm th}$ root of unity 
\begin{equation}
\alpha = \exp\left({2\pi i\over 6}\right).
\end{equation}
The usual choice for the periods in this basis appropriate for the polynomial M is $w^T=(w_2(\psi),w_1(\psi),w_0(\psi),w_5(\psi))$.
Then, the periods in the symplectic basis $\Pi(X)$ are related to the Picard--Fuchs periods by a linear transformation 
\begin{equation}
\Pi = \left( 
\begin{array}{c}
    {\cal F}_1 \\
    {\cal F}_2 \\
    X^1 \\
    X^2 \\
\end{array}
\right) = m\cdot \left( 
\begin{array}{c}
    w_2 \\
    w_1 \\
    w_0 \\
    w_5 \\
\end{array}
\right) 
\end{equation}
The matrix of transformation from the Picard--Fuchs to the symplectic basis has rational elements that are given by 
\begin{equation}
m=\left( 
\begin{array}{cccc}
    -{1 \over 3} & -{1 \over 3} & {1 \over 3} & {1 \over 3}\\
    0 & 0 & -1 & 0\\
    -1 & 0 & 3 & 2\\
    0 & 1 & -1 & 0\\
\end{array}
\right)\, .
\end{equation}
Quite generally, in the symplectic basis the periods have the expansion 
\begin{equation}
\Pi(\psi) = \sum_{k=1}^{\infty} c_k p_k \psi^{k-1}  \ ,
\end{equation}
where the $c_k$ are complex constants and $p_k$ are constant four-vectors with components that are rational linear combinations of powers of $\alpha$.

As a further computational tools, we note in passing that for this model the Meijer periods (we find them following method developed in \cite{Denef:2001xn} for quintic in the context of black hole attractors) are given by the following expression in Mathematica notations 
\begin{eqnarray}
&U^{-}_{0}(z)& = {\rm MeijerG}\left[\left\{\left\{\frac16,\frac26,\frac46,\frac56\right\},\{\}\right\},\{\{0\},\{0,0,0\}\},-z\right],\\
&U^{-}_{1}(z)& = {\rm MeijerG}\left[\left\{\left\{\frac16,\frac26,\frac46,\frac56\right\},\{\}\right\},\{\{0,0\},\{0,0\}\},z\right],\\
&U^{-}_{2}(z)& = {\rm MeijerG}\left[\left\{\left\{\frac16,\frac26,\frac46,\frac56\right\},\{\}\right\},\{\{0,0,0\},\{0\}\},-z\right],\\
&U^{-}_{3}(z)& = {\rm MeijerG}\left[\left\{\left\{\frac16,\frac26,\frac46,\frac56\right\},\{\}\right\},\{\{0,0,0,0\},\{\}\},z\right], 
\end{eqnarray}
where $U^{-}_{i}$ are valid for Im$(z) \leq 0$.

The transformation to the Picard--Fuchs basis is given by the matrix 
\begin{equation}
L=\left( 
\begin{array}{cccc}
    3 & -5 & 3 & -6\\
    -1 & 4 & 0 & 3 \\
    -1 & 0 & 0 & 0 \\
    3 & -4 & 3 & -3 \\
\end{array}
\right) 
\end{equation}
so that 
\begin{equation}
\left( 
\begin{array}{c}
    w_2(\psi) \\
    w_1(\psi) \\
    w_0(\psi) \\
    w_5(\psi) \\
\end{array}
\right) = L\cdot \left( 
\begin{array}{c}
    U^{-}_{0}(\frac{1}{\psi^6})\\
    U^{-}_{1}(\frac{1}{\psi^6})\\
    U^{-}_{2}(\frac{1}{\psi^6})\\
    U^{-}_{3}(\frac{1}{\psi^6})\\
\end{array}
\right) , 
\end{equation}
with $U^{-}\left(\frac{1}{\psi^6}\right)$ for Im$\left(\frac{1}{\psi^6}\right) \leq 0$.

\subsubsection{Two vacua for one flux and an interpolating domain wall} 

We now proceed to specifying the vacua for given values of the fluxes.
Let us take, as a convenient choice among the solutions of the F-flatness conditions, the flux components to be 
\begin{equation}
h = \{-50, -11, 6, -4\} 
\end{equation}
and 
\begin{equation}
f = - h \cdot A^2 = \{-4, -7, 7, 16\} .
\end{equation}
The virtue of this set of numbers is that the consequent vacua fall well within the convergence region of the Picard--Fuchs periods, thus making the numerical computations for the flow somewhat easier.

It turns out that these fluxes give rise to two vacua: the first vacuum is a maximum at the Landau--Ginzburg point, with 
\begin{equation}
\psi_1 = 0, \qquad \tau_1 = \frac{1+ i\sqrt{3}}{2}\, .
\end{equation}
The critical value of the potential is $V(\psi_1, \tau_1)=-3 e ^K|W|^2 \simeq -1219.5$.
We find the following diagonalized mass matrix for this vacuum 
\begin{equation}
{\rm mass} = {\rm diag}(-494.6, -494.6, -172.0, -172.0) .
\end{equation}
The second vacuum is a saddle point close to the conifold point, with 
\begin{equation}
\psi_2 \simeq 0.953 + 0.025 i, \qquad \tau_2 \simeq 0.162 + 1.088 i ,
\end{equation}
and with potential equal to $V(\psi_2, \tau_2) \simeq -1269$.
In this case the diagonalized mass matrix is 
\begin{equation}
{\rm mass} = {\rm diag}(1260.5, -377.6, -341.3, 171.3) .
\end{equation}

The interpolating solution going from one vacuum to the other is found by solving the flow equations (\ref{renorm1}), that become 
\begin{eqnarray}
&&\frac{d\psi}{du} = - g^{\psi\bar{\psi}} 
\partial_{\bar{\psi}}(K + \ln(\bar{W})) \ ,\\
&&\frac{d\tau}{du} = - g^{\tau\bar{\tau}} 
\partial_{\bar{\tau}}(K + \ln(\bar{W})) \ , 
\end{eqnarray}
where we have used for convenience the variable $u = \ln(a)$.

\begin{figure}[h!]
\centering 
\includegraphics[scale=.6]{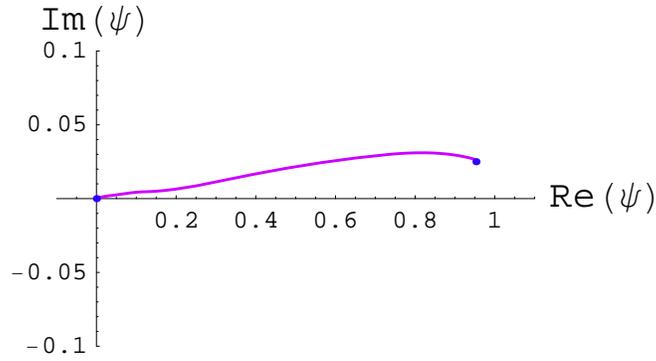} \caption{\small Flow on $\psi$ complex plane.}\label{plot1a} 
\end{figure}

\begin{figure}[h!]
\centering 
\includegraphics[scale=.6]{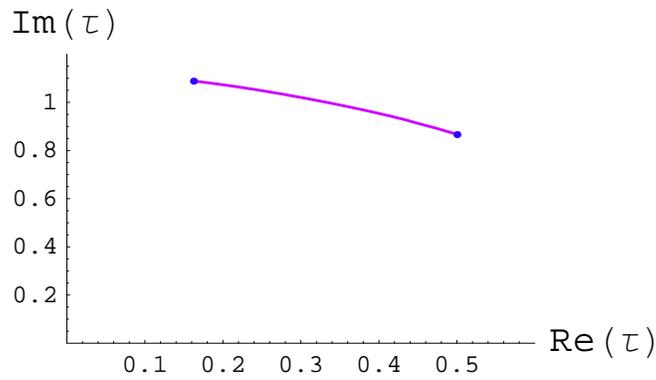} \caption{\small Flow on $\tau$ complex plane.}\label{plot2} 
\end{figure}

\begin{figure}[h!]
\centering 
\includegraphics[scale=.6]{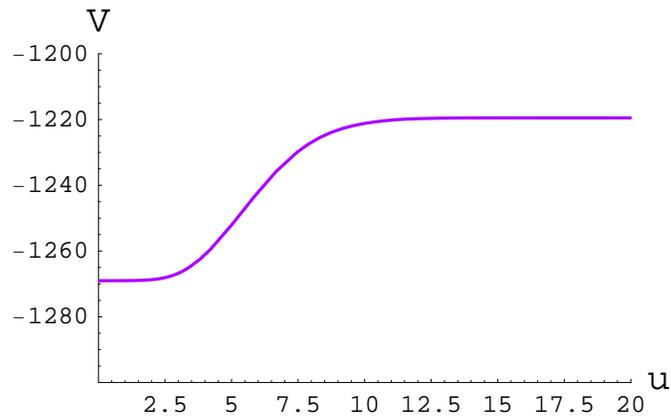} \caption{\small Flow of the potential $V(\psi(u),\tau(u))$.}\label{plot3} 
\end{figure}

The resulting flows are displayed below on the $\psi$ complex plane (Figure \ref{plot1a}) and on the $\tau$ complex plane (Figure \ref{plot2}).
The corresponding flow of the potential $V(\psi,\tau)$ as a function of $u= \ln (a)$ is shown in Fig.~\ref{plot3}.
It shows clearly that there is no barrier.
These domain walls are associated to flows that go from a maximum in the Landau--Ginzburg point, towards the conifold which is a saddle point.
These are explicit examples of critical points which may be sometimes repellers (for saddle points or maxima of the potential) and sometimes attractors, in particular stable attractors in case of minima of the potential.
In this case, the flow tends to escape from the LG point towards the conifold.

We are confident that these results on interpolating domain walls are common to other one-parametric Calabi--Yau models that have been studied in \cite{DeWolfe:2004ns} and generically have similar distribution of vacua on moduli space.

\section{BPS domain walls between vacua separated by barriers}

When the vacua are separated by a barrier, one can expect to find an absolute minimum as well as a meta-stable local minimum.
This class of models in the stringy landscape was suggested in \cite{Kallosh:2004yh}: in addition to the AdS minimum and to the asymptotically Minkowski vacuum, there is also a supersymmetric Minkowski minimum at some finite value of the volume modulus.
The main motivation for the new class of models with such a minimum was to have more flexibility with the gravitino mass, and to avoid certain cosmological problems specific to the standard KKLT scenario.
Here we will show that the same class of models (which we will call KL models) allows two AdS minima, both at finite value of the volume modulus.
Moreover, we will find two types of flows here: depending on the choice of parameters the superpotential may never vanish between the two AdS vacua, or it may vanish at some point and therefore change the sign along the flow.
This results in several different types of BPS domain walls in the KL models.

\subsection{KL model with Minkowski and AdS minima}\label{sec:KLmodel} 

\begin{figure}
[h!] \centering 
\includegraphics[scale=0.9]{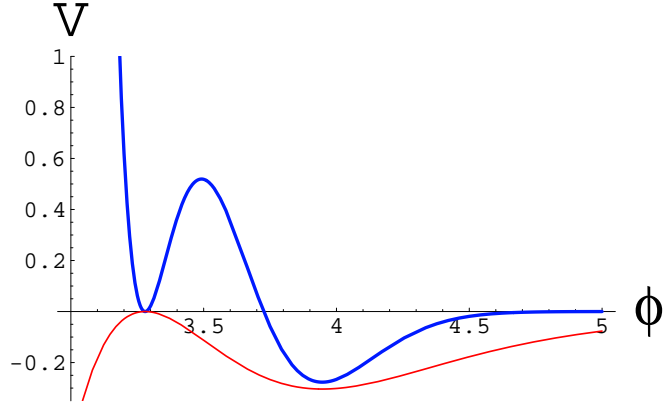} \caption{\small \small The potential energy density $V$, blue upper curve, and the covariantly holomorphic superpotential ${\cal Z}$, the red curve below, as the functions of the modulus $\phi$ for the KL model.
The potential is shown in units $10^{-8}$, and ${\cal Z}$ is in units $10^{-4}$.
Both $V$ and ${\cal Z}$ have extrema at the same point, which is a reflection of the fact that supersymmetry is unbroken in the minima of $V$.
This plot is for $A = 1$, $B = -2$, $W_0 = -0.125$, $a = \pi/100$ and $b= \pi/50$.}\label{KLPot} 
\end{figure}

\begin{figure}[h!]
\centering 
\includegraphics[scale=0.7]{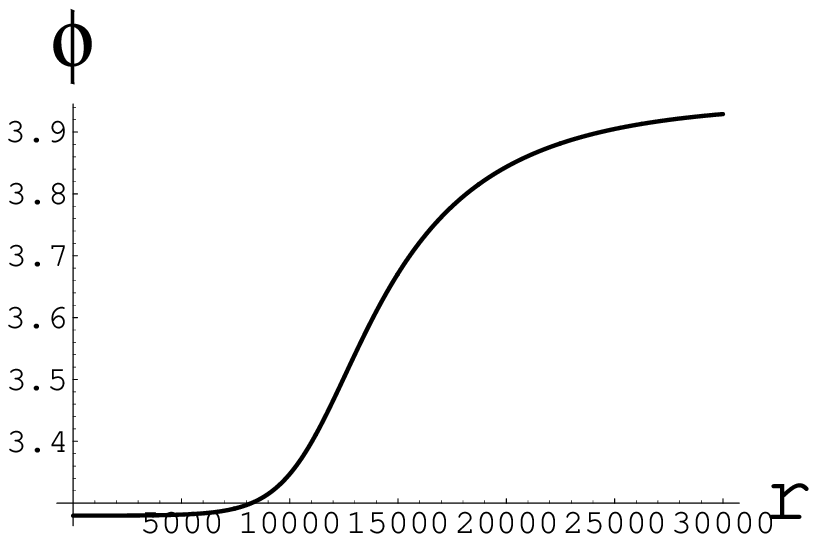}\hskip 0.5cm 
\includegraphics[scale=0.7]{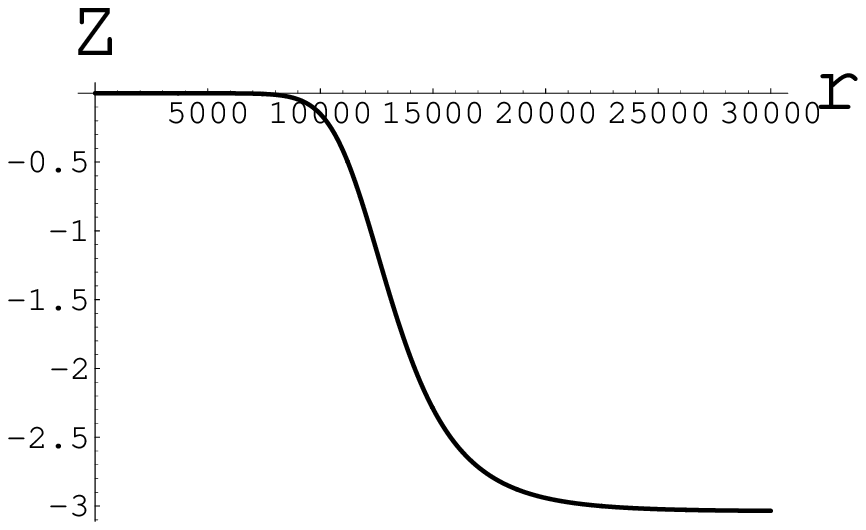} \vskip 0.7cm 
\includegraphics[scale=0.7]{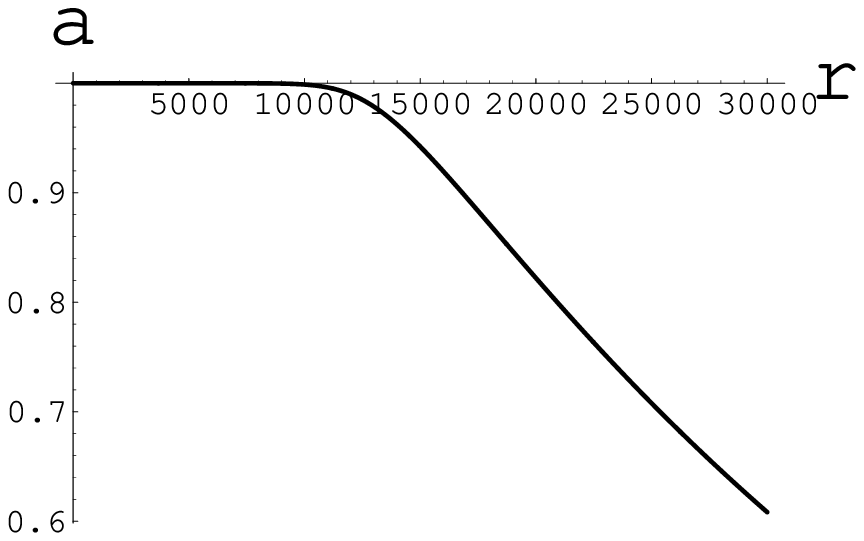}\hskip 0.5cm 
\includegraphics[scale=0.7]{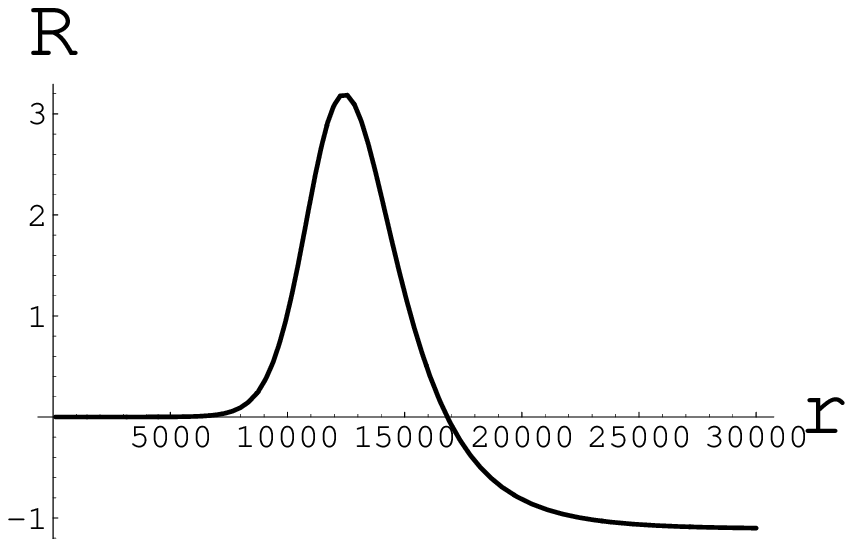} \caption{\small KL wall between M$_1$ and AdS.
We plot the scalar $\phi$, ${\cal Z}$ (in units $10^{-5}$), warp factor $a$ and the curvature scalar $R$ (in units $10^{-8}$) as functions of the coordinate $r$.
The curvature $R$ is equal to zero in the Minkowski vacuum, has a peak at the domain wall, and approaches a constant negative value in the AdS vacuum.}\label{plotx} 
\end{figure}

The model has one modulus as in the KKLT case, but a more involved superpotential of the racetrack type \cite{Kallosh:2004yh}.
The superpotential is a modification of the KKLT model by an additional exponential term 
\begin{equation}
W = W_0 + A e^{-a \rho}+ B e^{-b \rho} \ .
\end{equation}
This model is a simple instance of multiple basins of attraction.
It has one supersymmetric Minkowski vacuum at fixed volume, one supersymmetric AdS vacuum also at fixed volume and one supersymmetric Minkowski asymptotic vacuum at infinite volume as shown in an upper curve in Figure \ref{KLPot}.
The plot of the potential clearly shows the barrier between the left Minkowski minimum and the AdS minimum, the superpotential is negative everywhere away from Minkowski vacua.
For the choice $A = 1$, $B = -2$, $W_0 = -0.125$, $a = \pi/100$ and $b= \pi/50$, also using once more the definition $\rho = {\rm e}^{\frac{2}{\sqrt3} \phi}+ i \alpha$, the vacua are at $\alpha = 0$, $\phi_{M_1} \sim 3.3$ and $\phi_{AdS} \sim 3.94$, with $V \sim -2.7 \cdot 10^{-9}$.
Therefore, one can construct two interpolating domain walls.
One is between the AdS minimum and the local Minkowski minimum, which we will call M$_1$.
Another wall is between the AdS minimum and the Minkowski extremum at infinity, M$_\infty$.
Here we study only the domain walls interpolating between M$_1$ and AdS, see Figure \ref{plotx}.
The features of the domain wall between M$_\infty$ and AdS minimum in this model will be the same as the ones studied in KKLT case in Section \ref{sec:KKLTmodel}.
In Figure \ref{plotx} we plot on the top left the scalar potential $V$, and on the right the covariantly holomorphic superpotential ${\cal Z}$, which vanishes at M$_1$ and stays negative all the way towards AdS.
The warp factor on the bottom left starts as a constant at M$_1$ and interpolates into an exponent of $r$ near the AdS.
Finally, the curvature jumps from zero at M$_1$ to some positive value and falls down to a negative constant at AdS.
It is interesting that the curvature of the wall grows significantly between zero at M$_1$ before becoming negative at AdS.

\subsection{The KL model with two different AdS minima}\label{adskl} 
\subsubsection{BPS walls with critical tension}\label{critsect} 

\begin{figure}
[h!] \centering 
\includegraphics[scale=0.9]{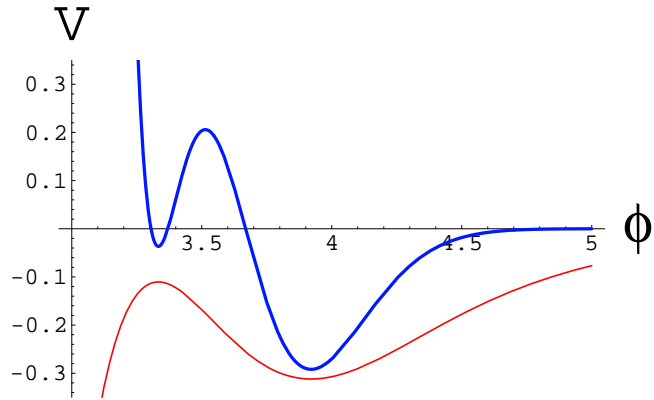} \caption{\small The potential $V$ (upper curve, in units $10^{-8}$) and the covariantly holomorphic superpotential ${\cal Z}$ (lower curve, in units $10^{-4}$) for the KL model, as function of the modulus.
Notice that we took $A = 0.96$ instead of $A=1$, all other parameters being the same as for the potential in Fig.~\ref{KLPot}.}\label{fig:KLPpota} 
\end{figure}

Now we are going to consider the models with two AdS minima.
This can be done, e.g., by taking the KL model with a supersymmetric Minkowski minimum M$_{1}$, and change any of the parameters of this model.
In particular, as one can see from Fig.\ref{fig:KLPpota}, one can obtain an AdS minimum from the Minkowski minimum by a slight modification of a single parameter in the previous model: we changed $A=1$ to $A=0.96$.

Using once more the definition $\rho = {\rm e}^{\frac{2}{\sqrt3} \phi}+ i \alpha$, the AdS vacua are at $\alpha = 0$, $\phi_{0} \sim 3.3$ and $\phi_{1} \sim 3.9$, with $V_{{{0}}} \sim -0.034 \cdot 10^{-8}$, and $V_{{{1}}} \sim -0.29 \cdot 10^{-8}$ for $\alpha = 0$.
The potential $V$ (upper curve) and the covariantly holomorphic superpotential ${\cal Z}$ (lower curve) for the KL model as function of the modulus are shown in Figure \ref{fig:KLPpota}, for the choice $A = 0.96$, $B = -2$, $W_0 = -0.125$, $a = \pi/100$ and $b= \pi/50$.
${\cal Z}$ is always negative.

\begin{figure}[h!]
\centering 
\includegraphics[scale=0.75]{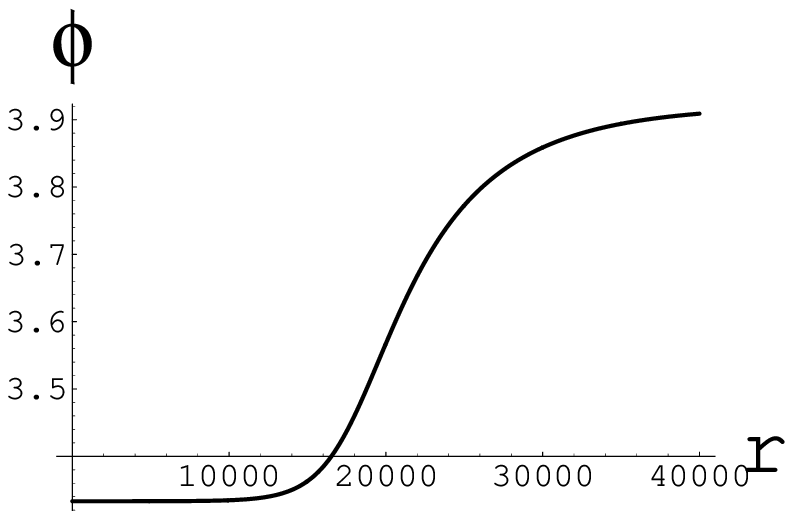}\hskip 0.5cm 
\includegraphics[scale=0.75]{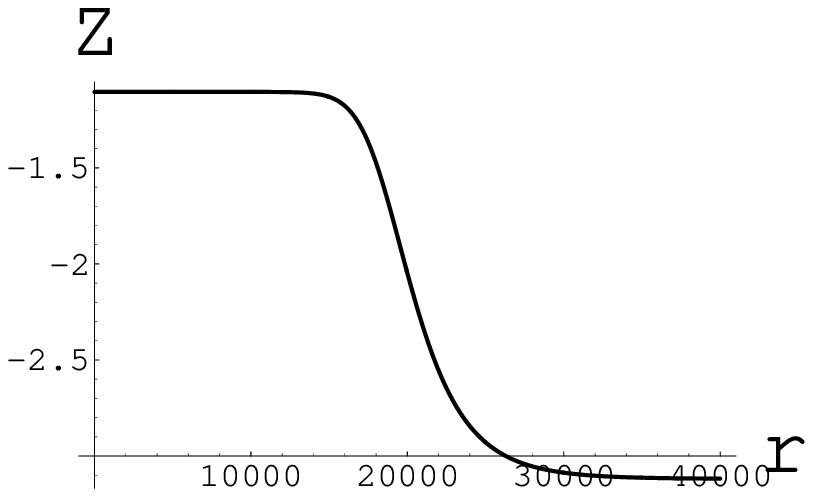} \vskip 0.7cm 
\includegraphics[scale=0.75]{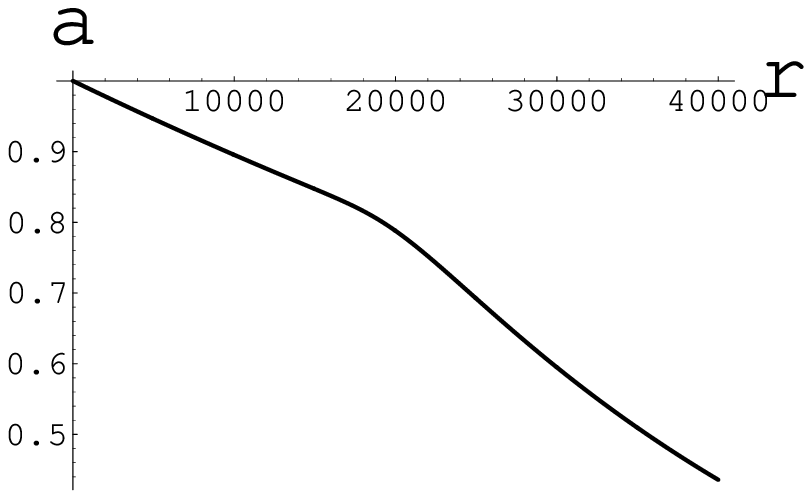}\hskip 0.5cm 
\includegraphics[scale=0.75]{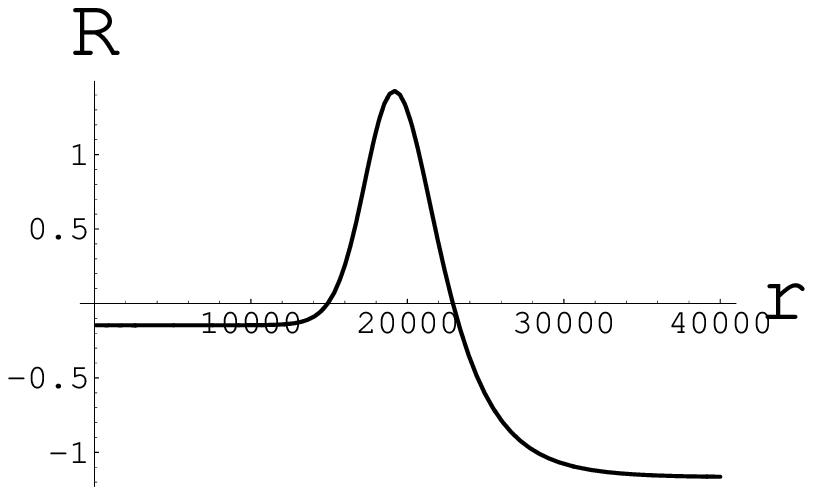} \caption{\small KL wall with a critical tension between $AdS_1$ and $AdS_{2}$.
We plot the scalar $\phi$ and the covariantly holomorphic superpotential ${\cal Z}$ (in units $10^{-5}$), which is always negative, warp factor $a$, and the curvature scalar $R$ (in units $10^{-8}$), as a function of the coordinate $r$.
The curvature $R$ has a peak at the domain wall, and approaches different constant negative values in two different AdS minima.} \label{fig:plota} 
\end{figure}

We show various features of the KL wall between the two AdS vacua in Fig.~\ref{fig:plota}.
We plot the scalar and the covariantly holomorphic superpotential, which is a negative constant near the first AdS vacuum and interpolates towards another negative value when it approaches another AdS vacuum.
The warp factor $a$ starts as a decreasing exponent and then changes into a faster decreasing exponent of $r$ near the second AdS.
The curvature $R$ jumps from some constant negative value near the first AdS vacuum, becomes positive near the domain wall, and then falls down to a negative constant near the second AdS vacuum.

The tension of this BPS wall is given by Eq.~(\ref{tension1}).
\begin{equation}
\sigma = {2\over \sqrt 3} (|V_{1}|^{1/2} - |V_{0}|^{1/2}) \ .
\label {supercrit} 
\end{equation}
It is critical in a sense explained in \cite{Cvetic:1996vr}: The tension of such BPS brane corresponds to a limit of the CDL bubble solution \cite{Coleman:1980aw} with infinite radius.
When such BPS wall exist there is no tunneling since these supersymmetric walls saturate the CDL bound.
We will discuss this issue later when we will study tunneling, see Section \ref{decay}.

\subsubsection{BPS walls with super-critical tension}\label{sec:subcritical}

In the previous subsection we considered a KL model with two AdS minima and the superpotential which did not change sign on the way from one minimum to another.
Now we will construct a KL model with two AdS minima and superpotential vanishing at some point between them  (Fig. 12).
This can be done by a small deviation of one of the parameters: instead of taking $A=1$ as in original KL model, or $A<1$, as in the previous subsection with two AdS vacua, it is sufficient to take $A>1$.

For example, for the choice $A = 1.05$, $B = -2$, $W_0 = -0.125$, $a = \pi/100$ and $b= \pi/50$, also using once more the definition $\rho = {\rm e}^{\frac{2}{\sqrt3} \phi}+ i \alpha$, one finds that the AdS vacua are at $\alpha = 0$, $\phi_{0} \sim 3.2$, with $V_{0} \sim -0.8 \cdot 10^{-9}$, and at $\alpha = 0$, $\phi_{1} \sim 4$, with $V_{1} \sim -2.6 \cdot 10^{-9}$.
The covariantly holomorphic superpotential ${\cal Z}$ is positive near the AdS minimum at $\phi_{0}$ .
It crosses zero at $\phi \sim 3.45$ and becomes negative when approaching the second AdS minimum at $\phi_{1} \sim 4$.
It approaches zero at infinite values of the volume modulus.
\begin{figure}
[h!] \centering 
\includegraphics[scale=0.9]{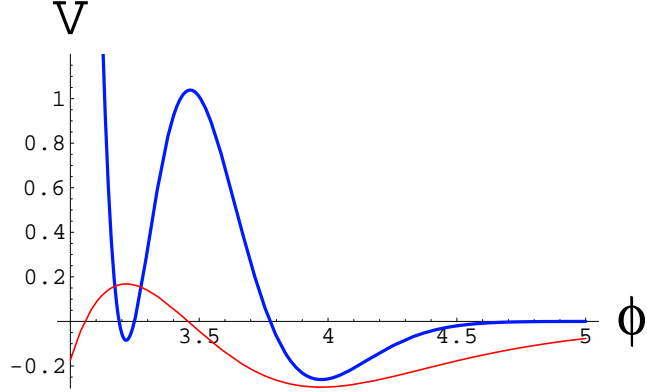} \caption{\small The potential $V$ (upper curve, in units $10^{{-8}}$) and the covariantly holomorphic superpotential ${\cal Z}$ (lower curve, in units $10^{{-4}}$) for the KL model as function of the modulus.
${\cal Z}$ is positive at $\phi_{0}$ and negative at $\phi_{1}$.
Here we took $A = 1.05$ instead of $A=1$, all other parameters being the same as for the potential in Fig.~\ref{KLPot}.}\label{KLPotb} 
\end{figure}

\begin{figure}[h!]
\centering 
\includegraphics[scale=0.75]{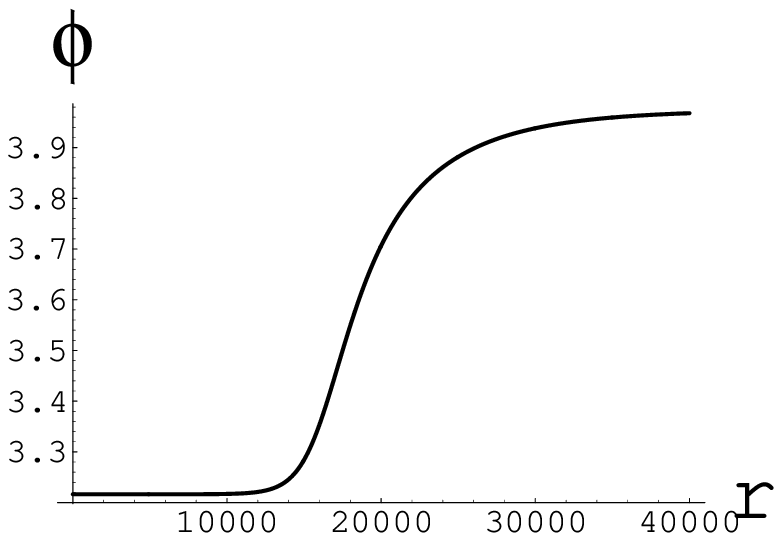}\hskip 0.5cm 
\includegraphics[scale=0.75]{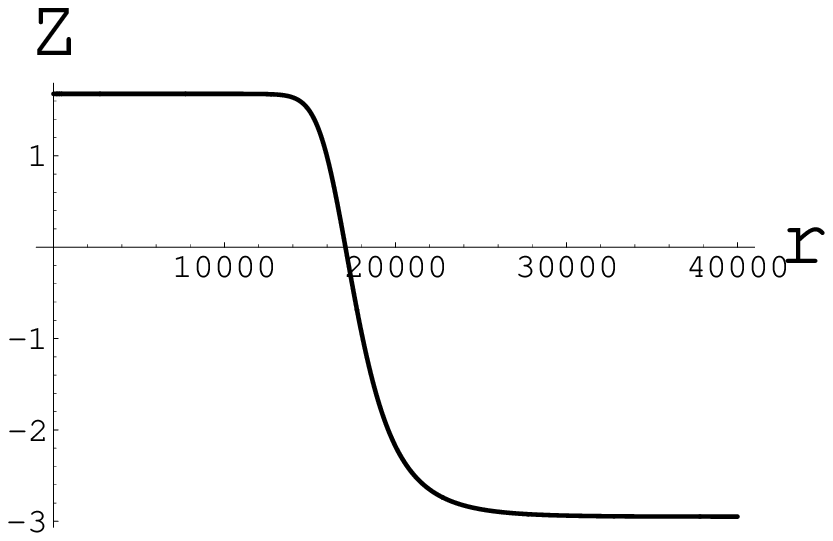} \vskip 0.3cm 
\includegraphics[scale=0.75]{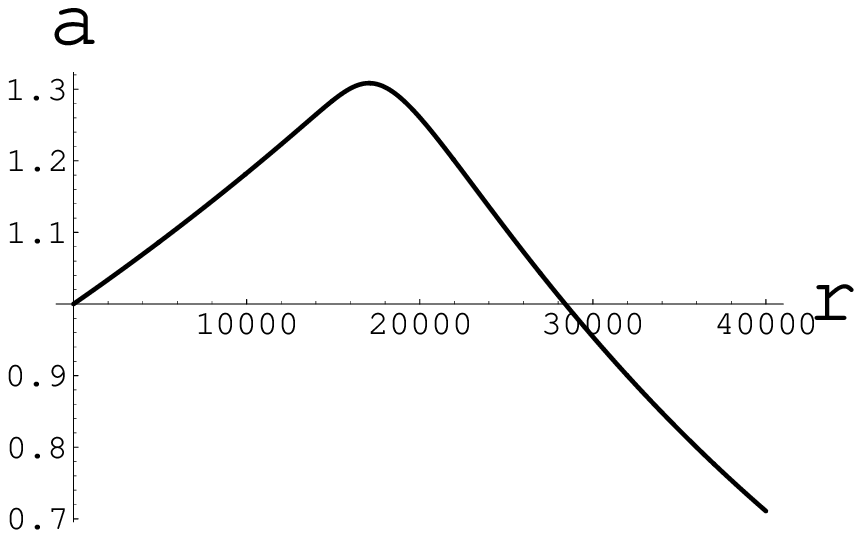} 
\includegraphics[scale=0.75]{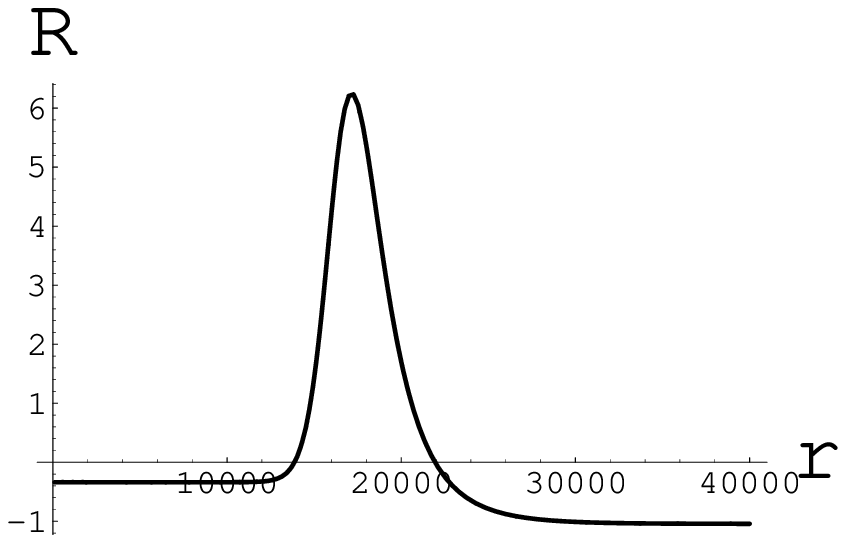} \caption{\small A BPS wall between two AdS domains in the version of the KL model where ${\cal Z}$ changes the sign on the domain wall solution.
We plot the scalar $\phi$, the covariantly holomorphic superpotential ${\cal Z}$ (in units $10^{-5}$), the warp factor $a$, and the curvature $R$ (in units $10^{-8}$).
Notice the characteristic peak of the scale factor at the point where ${\cal Z}$ changes sign.}\label{plotb} 
\end{figure}

The corresponding BPS solution is shown in Fig.~\ref{plotb}.
We plot on the top left the scalar field $\phi$, and on the right the covariantly holomorphic superpotential, which changes the sign on the solution.
The warp factor on the bottom left starts as a growing exponent near the first AdS minimum and changes into a decreasing exponent of $r$ near the second minimum.
Finally, the curvature jumps from some constant negative value near the first AdS vacuum, becomes positive near the domain wall, and then falls down to a negative constant near the second AdS vacuum.
As far as we know, this solution provides the first explicit example of a smooth BPS domain walls in $d=4$ with the superpotential vanishing between the critical points.

The tension of this BPS wall is given by Eq.~(\ref{tension2}).
\begin{equation}
\sigma = {2\over \sqrt 3} (|V_{1}|^{1/2} + |V_{0}|^{1/2}) \ .
\label {crit} 
\end{equation}
It is super-critical in the sense explained in \cite{Cvetic:1996vr}.
The tension of such BPS walls exceeds the CDL bound and therefore the tunneling is super-suppressed.

\section{Domain walls, near-BPS bubbles, and vacuum decay}\label{decay} 
\subsection{Decay of an uplifted vacuum to a collapsing universe}

In previous sections we studied domains with unbroken supersymmetry separated by static BPS domain walls.
In our world supersymmetry is broken, but if supersymmetry breaking is not very large as compared to other parameters, one may hope that our domain wall solutions will provide an approximate description of slowly moving near-BPS domain walls.
Just as in the situation with the near-extremal black holes, one may hope the physics of the near-BPS walls can be under a much better theoretical control than the physics of generic domain walls.
It would be especially interesting to see whether the BPS domain walls may play any role in the cosmological context, dividing the universe into different regions separated by static domain walls.
However, from the point of view of the Friedmann cosmology, the universe with a negative cosmological constant $\Lambda <0$ should collapse to a cosmological singularity within a very short time $t \sim |\Lambda |^{{-1/2}}$.
Therefore in the usual cosmological context one can hardly describe coexisting AdS domains separated by static domain walls.

One of the most interesting applications of the results obtained in the last section is the theory of vacuum decay and bubble formation.
It is well known that dS vacua obtained in the KKLT construction by uplifting of AdS vacua are unstable with respect to the decay to the Minkowski vacuum corresponding to infinitely large values of the volume modulus $\phi$, i.e.
to a supersymmetric 10D vacuum \cite{Kachru:2003aw}.

According to \cite{Coleman:1980aw}, the decay probability is given by 
\begin{equation}
\label{prob} P(\phi) = e^{-B} = e^{-S(\phi)+S_0}, 
\end{equation}
where $S(\phi)$ is the Euclidean action for the tunneling trajectory, and $S_0=S(\phi_0)$ is the Euclidean action for the initial metastable configuration $\phi = \phi_0$.

The Euclidean action calculated for the metastable vacuum dS solution $\phi=\phi_0$ is given by 
\begin{equation}
\label{action2} S_0 = - {24\pi^2\over V_0} <0\ .
\end{equation}
This action has a simple sign-reversal relation to the entropy of de Sitter space ${\bf S_0}$: 
\begin{equation}
\label{action2a} {\bf S_0} = - S_0 =+ {24\pi^2\over V_0}\ .
\end{equation}
Therefore the decay time of the metastable dS vacuum  $t_{\rm decay} \sim P^{-1}(\phi)$ can be represented in the following way: 
\begin{equation}
\label{decaytime} t_{\rm decay} = e^{S(\phi)+\bf S_0} = t_r \ e^{S(\phi)}\ .
\end{equation}
Here $t_{r} \sim e^{\bf S_{0}}$ is the so-called recurrence time.
It was shown in \cite{Kachru:2003aw} that if the decay of the metastable dS vacua occurs due to the tunneling through the barrier with a positive scalar potential, then $S(\phi)$ is always negative, and therefore the decay always happens during the time shorter than the recurrence time $t_{r}$.
This result directly applied to the simplest KKLT model where the tunneling occurs through the positive barrier separating the metastable dS vacuum and the supersymmetric 10D vacuum.
However, the situation with the tunneling to the minima with a negative vacuum density remained less clear; see Refs.
\cite{Banks:2005ru,Aguirre:2006ap,Bousso:2006am} for the recent discussion.
In what follows we will examine this issue using the results obtained in the previous section.

Our consideration will be based on the observation of Ref.~\cite{Cvetic:1996vr} that in all cases when $Z$ does not vanish across the domain wall, the domain wall solutions can be represented as a limiting configuration describing the wall of the CDL bubbles of an infinitely large radius.
For such bubbles, the tunneling action is also infinitely large, and the vacuum decay is impossible.
In our context, this fact is related to the supersymmetry of the different vacua \cite{Weinberg:1982id}, and of the interpolating BPS wall solution.

However, in realistic situations supersymmetry is broken.
For example, in the KKLT construction the metastable dS vacuum state appears after the uplifting of a supersymmetric AdS vacuum.
Uplifting can be achieved, e.g., by adding a $\overline{D3}$ brane contribution, which results in a supersymmetry breaking in the dS vacuum.
Similarly, in the KL model \cite{Kallosh:2004yh} one must uplift a Minkowski vacuum or an AdS vacuum to a dS vacuum, thus breaking the supersymmetry.
As we will see, this may lead to a relatively rapid decay of the uplifted dS vacuum due to creation of the bubbles describing a collapsing universe with negative vacuum energy corresponding to the AdS vacuum.

Note that in terms of a canonically normalized volume modulus $\phi$, which we will use hereafter, the $\overline{D3}$ brane contribution looks as $C e^{-2\sqrt2 \phi/\sqrt3}$, where $C$ is some constant.
(The canonically normalized field, $1/2 ( 
\partial \phi)^2$, is obtained by dividing the field $\phi$ used in the previous section by the factor $\sqrt 2$.) This contribution rapidly decreases at large $\phi$.
Therefore this term may significantly uplift the AdS (or Minkowski) minimum at small values of $\phi$, but it makes a much smaller effect on the AdS minima at large $\phi$.
This suggests that the transitions should typically occur towards larger values of the volume of the compactified space.

\subsubsection{Decay of an uplifted Minkowski vacuum}\label{decmink}

First of all, consider the KL potential with the supersymmetric Minkowski vacuum, and uplift it by a tiny amount to make the vacuum energy there equal to the present cosmological constant, $V_{0} \sim 10^{{-120}}$, in Planck energy density units.
This should make this dS vacuum metastable, and lead to a vacuum decay due to creation of bubbles with negative vacuum energy.
\begin{figure}[hbt] 
\centering 
\includegraphics[scale=0.75]{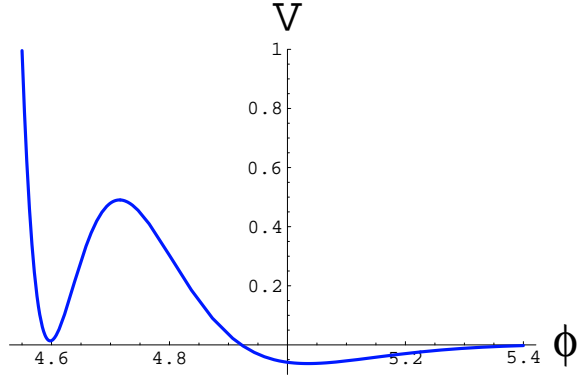} \caption{\small A slightly uplifted KL potential, shown in units $10^{{-12}}$.}\label{KLthin} 
\end{figure}
To simplify the investigation of its decay, we will consider potentials such that the difference between the depth of the two minima is much smaller than the height of the barrier.
An example of such a potential is provided by the KL model with parameters $A = 0.1$, $B = -10$, $W_0 = -0.00025$, $a = 2\pi/50$ and $b= 2\pi/25$.
For these values of parameters the potential has a supersymmetric Minkowski minimum at $\phi \sim 4.6$ and an AdS minimum at $\phi \sim 5$.
If one slightly changes the parameters, e.g.
takes $ A = 0.095$, then the Minkowski minimum becomes an AdS minimum.
If one uplifts the potential by adding a term $C e^{-2\sqrt2 \phi/\sqrt3}$, the potential acquires a dS minimum, as shown in Fig.~\ref{KLthin}.
During all of these changes, the second AdS minimum changes only slightly.

One expects that the thin wall approximation should be valid for the description of tunneling in such potentials.
One part of this condition is automatically satisfied for the near-BPS bubbles.
Indeed, by considering sufficiently small uplifting one can always make such bubbles arbitrarily large without changing the thickness of the wall.
However, in order to use the thin-wall approximation as formulated in Ref.~\cite{Coleman:1980aw}, one should also check that the difference between the depth of the two minima is much smaller than the height of the barrier, and that the width of the wall of the CDL bubble remains much smaller than the radius of the bubble even if one turns off the gravitational interaction.
We checked numerically that this condition is also satisfied for our potential.

Having in mind potentials of this type, we will use the thin-wall approximation to investigate the decay of the dS vacuum with the field $\phi_{0}$ and vacuum energy density $V_{0} = V(\phi_{0}) > 0$ to the state with the field $\phi_{1}$ with a negative vacuum energy density $V_{1} = V(\phi_{0}) < 0$.
We will be interested in the realistic situation $|V_{0}| \ll |V_{1}|$, as illustrated by Fig.~\ref{KLthin}.
Examples of the vacuum decay considered in \cite{Coleman:1980aw} described transitions between dS and Minkowski space, and decay of the Minkowski vacuum.
We will use the results obtained by Parke \cite{Parke:1982pm}, who generalized the results of \cite{Coleman:1980aw} for the tunneling between the states with arbitrary values of $V_{i}$ in the thin-wall approximation.

According to \cite{Parke:1982pm}, the bubble size is given by 
\begin{equation}
\label{rad} \rho^{2} = {\rho_{0}^{2}\over 1 + 2{\rho_{0}^{2}\over 4\lambda^{2}}+\left(\rho_{0}\over 2\Lambda\right)^{4}} \ , 
\end{equation}
where  
\begin{equation}
\rho_{0} = {3\sigma\over V_{0}-V_{1}} 
\end{equation}
is the bubble radius in the thin wall approximation ignoring gravity, $\sigma$ is the bubble wall tension, and 
\begin{equation}
\lambda^{2}= {3\over V_{0}+V_{1}} 
\end{equation}
(it can be positive or negative; in our case it is negative), and 
\begin{equation}
\Lambda^{2}= {3\over V_{0}-V_{1}} \ .
\end{equation}
In the thin-wall approximation, the tension of the wall can be approximated by the integral 
\begin{equation}
\label{tension} \sigma = \int_{{\phi}_{0}}^{{{\phi}^{*}_{1}}}[V(\phi)-V(\phi_{0})]^{1/2} \ .
\end{equation}
Here $\phi*_{1}$ is the point near $\phi_{1}$ where $V(\phi)-V(\phi_{0}) =0$.
(Ref.~\cite{Coleman:1980aw} uses a slightly different prescription, which is equivalent to ours in the limit when the height of the barrier is much greater than the difference between $V_{i}$.) 
In the limit when the radius of the CDL bubble becomes infinitely large, the wall tension is known exactly: 
\begin{equation}
\label{tension2aa} \sigma = {2\over \sqrt 3} (|V_{1}|^{1/2} - |V_{0}|^{1/2}) \ , 
\end{equation}
see \cite{Cvetic:1996vr,Parke:1982pm}, and also our derivation of this result in Sect.
2, Eq.~(\ref{tension1}).

Using the variables 
\begin{equation}
x = \left({\rho_{0}\over 2\Lambda}\right)^{2} , \qquad y = {\Lambda^{2}\over \lambda^{2}} ,
\end{equation}
one can represent Eq.~(\ref{rad}) in the following convenient form: 
\begin{equation}
\label{size} \rho^{2} = {\rho_{0}^{2}\over 1 + 2xy+x^{2}} \ .
\end{equation}
For a particular case of tunneling from Minkowski space ($V_{0} = 0$), the infinite bubble size regime (the threshold regime for the tunneling) corresponds to $x = 1$, $y = -1$, when the denominator vanishes.

The tunneling probability is given by $e^{{-B}}$, where  
\begin{equation}
\label{coeffB} B= {27 \pi^2\sigma^{4}\over 2(V_{0}-V_{1})^{3}} ~ r(x,y) \ .
\end{equation}
Here the first term is the no-gravity result, and the gravitational correction factor $r(x,y)$ is given by \cite{Parke:1982pm} 
\begin{equation}
r(x,y) = \frac{2[(1+x y)-(1+2xy+x^2)^{\frac{1}{2}}]}{x^2(y^2-1)(1+2xy+x^2)^{\frac{1}{2}}} \ .
\label{gravfactor} 
\end{equation}
This factor also blows up at $x = 1$, $y = -1$, i.e.
the tunneling becomes forbidden.

Now let us uplift the Minkowski vacuum, assuming that the AdS vacuum at $\phi_{1}$ remains approximately at the same height $V_{1}$.
As we already argued, this is indeed the case if the AdS vacuum is at large values of the volume modulus $\phi$.
Uplifting changes not only $V_{0}$, but also the bubble wall tension $\sigma$.
Typically uplifting decreases the wall tension, which increases the probability of the tunneling; a sufficiently large uplifting may remove the wall altogether, see \cite{Kallosh:2004yh}.

Any small uplifting makes the tunneling from $\phi_{0}$ to $\phi_{1}$ possible.
In order to study the size of the bubbles and the probability of the tunneling, let us change $x$ and $y$ by $\Delta x$ and $\Delta y$.
In the linear approximation $\Delta x\sim \Delta y$ and both depend on the small uplifting $ V_0 \ll 1$.
In general, one would expect that the result should depend both on $\Delta x$ and $\Delta y$.
However, we find that in the leading approximation in $\Delta x$ and $\Delta y$, the results of the expansion near $x = 1$, $y = -1$ depend only on $\Delta y$, i.e.
they are not sensitive to the small changes of the bubble wall tension.
In particular, the denominator in (\ref{size}) is equal to $2\, \Delta y$, where $\Delta y = -{2V_{0}\over V_{1}}$, for $|V_{0}| \ll |V_{1}|$.
This yields 
\begin{equation}
\rho^{2} = {\rho_{0}^{2}\over 2 dy} = -{ 9 \sigma^{2} \over 4V_{0} V_{1} }= {3\over V_{0} } = H_{1}^{-2} \ .
\end{equation}
Here $H_{1}$ is the Hubble constant in dS space with vacuum energy density $ V_{0}$.
Thus, instead of an infinite size bubble, we have a bubble with the size equal to the size of the dS horizon!

Now we will perform a similar investigation for the tunneling rate.
By considering the function $r^{{-1}}(x,y)$ in Eq.~(\ref{gravfactor}) and expanding it with respect to $\Delta x$ and $\Delta y$, one finds that in the limit $|V_{0}| \ll |V_{1}|$ 
\begin{equation}
B= {12 \pi^{2} \over V_{0} } = {\bf S_{0}}/2 \ .
\end{equation}
where ${\bf S_{0}}$ is the entropy of dS space, $ {\bf S_{0}}= {24 \pi^{2} \over V_{0} }.$
This means that the tunneling probability is given by the universal model-independent equation which has a simple geometric interpretation in terms of dS entropy: 
\begin{equation}
P \sim e^{-{\bf S_{0}}/2} \ .
\end{equation}
Several comments are in order.
First of all, we have found that the decay rate of the vacuum obtained by uplifting of a supersymmetric Minkowski minimum does not depend on the tension of the wall, or on the depth of the supersymmetric AdS minimum.
The only thing we needed to know is that prior to the uplifting these two minima were supersymmetric, which is a generic property of all known flux vacua.
An unexpected feature of this result is that the tunneling is suppressed not by a factor $e^{-{\bf S_{0}}}$, but by a factor $e^{-{\bf S_{0}}/2}$.
This means that the decay time of an uplifted Minkowski vacuum will be shorter than the recurrence time $t_{r} \sim e^{{\bf S_{0}}}$ by a huge factor $e^{{\bf S_{0}}/2} \sim e^{10^{120}/2}$.
We will return to the discussion of this fact in Section \ref{sink}.

\subsubsection{Decay of an uplifted AdS vacuum}\label{decads}

As we have seen in the previous subsection, a tiny supersymmetry breaking achieved by uplifting of the supersymmetric Minkowski vacuum to the dS vacuum with $V_{0} \sim 10^{-120}$ renders this vacuum unstable.
Now we will extend our results for a more general case.

Suppose that initially we have two supersymmetric AdS vacua, with depth $V_{0}, V_{1} < 0$, such that $|V_{0}| < |V_{1}|$.
One can achieve it e.g.
by considering the KL potential discussed in the previous section, but taking $A = 0.09$ instead of $A = 0.095$.
Tunneling between the two supersymmetric AdS minima is forbidden.
Let us also assume that the deeper minimum is at larger values of the volume modulus, so the KKLT uplifting of the minimum $V_{0}$ to dS space with $V_{0} \sim 10^{-120}$ leaves the depth of the second AdS minimum almost unaffected.
This is indeed the case in the model discussed above.
We will also assume that prior to the uplifting $|V_{0}|$ was much greater than the present value of the vacuum energy $10^{-120}$, so for all practical purposes it is not important whether we uplifted it to a dS state $V_{0} \sim 10^{-120}$ or to the Minkowski state with $V_{0} = 0$.
This subtle difference was important in our previous investigation because the degree of supersymmetry breaking was proportional to $V_{0} \sim 10^{-120}$.
Now we are considering a much greater supersymmetry breaking, which is achieved due to a much more significant uplifting, from the initial value of $V_{0}$ to (approximately) zero.
In order to simplify notation and avoid mixing the values of $V_{0}$ before and after the uplifting, we will assume that after the uplifting $V_{0} = 0$, and in all subsequent equations we will understand by $V_{0}$ its value in the AdS minimum prior to the uplifting.

In order to study the decay rate of the (nearly) Minkowski vacuum formed by uplifting, one should use, as before, eqs.~(\ref{coeffB}) and (\ref{gravfactor}).
In Minkowski vacuum, one has $y = -1$, so the expression (\ref{gravfactor}) looks singular, but in fact this singularity is fictitious, which can be easily seen by expanding it near $y = -1$.

The outcome of the calculations can be represented in the following form: 
\begin{equation}
\label{tunnads} B = {27 \pi^{2}\sigma^{4}\over 2 |V_{1}|^{3} }\cdot \left(1 -{9\sigma^{4}\over 16 |V_{1}|^{2}}\right)^{{-1}}\ .
\end{equation}
Here  $\sigma$   is the bubble wall tension for the tunneling in the uplifted potential.  The first term in this expression is the  probability of the decay of the uplifted Minkowski vacuum ignoring gravitational corrections. The second term represents the  gravitational suppression of the probability of the tunneling due to the factor $r(x,y)$.

As we see, the probability of the tunneling depends on the wall tension $\sigma$.
However, the only part of the expression (\ref{tunnads}) where this dependence is crucially important is the gravitational suppression term $\bigl(1 -{9\sigma^{4}\over 16 |V_{1}|^{2}}\bigr)^{{-1}}$. If after uplifting we would still have two supersymmetric vacua (Minkowski and AdS), then we would have $\sigma = {2\over \sqrt 3} |V_{1}|^{1/2} $, the last term in (\ref{tunnads}) would blow up, and the tunneling would be forbidden.
However, prior to the uplifting we had a smaller value of tension, $\sigma = {2\over \sqrt 3} (|V_{1}|^{1/2} - |V_{0}|^{1/2})$.
As we argued before, uplifting tends to decrease the barrier between the two vacua, and therefore it tends to decrease the tension. (Our numerical investigation of the tunneling in the KL model with the potential shown in Fig.~\ref{KLthin} is consistent with this expectation.) If this is the case, the gravitational suppression gradually disappears, and the probability of the tunneling becomes very large.

To illustrate various options, let us assume, in the first approximation, that $\sigma$ does not change during the uplifting.
In this case we find that 
\begin{equation}
\label{tunndsads} B= {24 \pi^2\over |V_{0}|}\, {(C-1)^{4}\over C^{2}(2C-1)^{2}} ,
\end{equation}
where $C^{2} = {|V_{1}|\over |V_{0}|} \geq 1$.

For $C = O(1)$ (i.e. for $V_{1}$ of the same order as $V_{0}$, prior to the uplifting), this expression coincides, up to a numerical coefficient ${(C-1)^{4}\over C^{2}(2C-1)^{2}} = O(1)$, with the entropy of dS space with the vacuum energy $|V_{0}|$, see Eq.~(\ref{action2a}).
The main difference is that for the dS space in which we live now we have $V_{0} \sim 10^{-120}$, and the factor $e^{-\bf S_{0}}$ is incredibly small, $e^{-\bf S_{0}} \sim e^{-10^{120}}$.
Meanwhile the depth of the  AdS vacuum prior to the uplifting can be a hundred orders of magnitude greater than $10^{-120}$, and therefore the decay rate of this vacuum after the uplifting can be very large (see below for a numerical estimate).
Note that if, as we expect, $\sigma$ slightly decreases during the uplifting in our model, then $B$ will be even  smaller, and  the decay rate will be even higher.

It would be nice to confirm our conclusions by other methods, or to calculate the tunneling probability directly, going beyond the thin-wall approximation.
However, our main goal here was to show that generically there is no reason to expect a strong gravitational suppression of the decay of dS vacua if they are obtained by the KKLT uplifting of the supersymmetric AdS minima.
Qualitatively, the main argument is as follows: If there are many AdS minima in the stringy landscape, and if the tunneling between them is suppressed by supersymmetry, then this suppression disappears after the uplifting and supersymmetry breaking.

Let us check whether these intuitive expectations match our estimates, Eq.~(\ref{tunndsads}).
According to (\ref{potential2}), the scalar potential prior to the uplifting is given by $V=|D_i {\cal Z}|^{2} - 3 |{\cal Z}|^2$.
In the supersymmetric minimum at $\phi_{0}$ we have $D_i {\cal Z} = 0$, and $V_{0} = - 3 |{\cal Z}|^2$.
Uplifting practically does not change the position of the minimum $\phi_{0}$, and therefore it practically does not change ${\cal Z}(\phi_{0})$.
Meanwhile the gravitino mass squared after the uplifting is given by $m_{3/2}^{2} = |{\cal Z}|^2$. This means that the depth of the AdS minimum before the uplifting is equal to $-3 m_{3/2}^{2}$, where $m_{3/2}$ is the present value of the gravitino mass \cite{Kallosh:2004yh}.
Therefore our result for the tunneling rate of the uplifted AdS state (\ref{tunndsads}) can be represented in the form directly related to the gravitino mass: 
\begin{equation}
\label{tunndsadsgrav} B= {8 \pi^2\over m_{3/2}^{2}}\, {(C-1)^{4}\over C^{2}(2C-1)^{2}}  \ .
\end{equation}

The scale of supersymmetry breaking in our world is much greater than the energy scale corresponding to the present value of the cosmological constant.
For example, if the gravitino mass is $m_{3/2} \gtrsim 10^{2}$ GeV $\sim 10^{{-16}}$ in Planck units, then $m_{3/2}^{2} \gtrsim 10^{{-32}}$.
Therefore Eq.~(\ref{tunndsadsgrav}) suggests that a typical decay rate of the AdS vacua after their uplifting to the dS state with $V_{0} \sim 10^{-120}$ is greater than $e^{-10^{34}}$.
Most importantly, this decay rate is greater than $e^{-\bf S_{0}} \sim e^{-10^{120}}$ by an enormously large factor $e^{-10^{34}}/e^{-10^{120}} =  e^{(10^{120}-10^{34})} \approx e^{10^{120}}$. 

This conclusion may have important cosmological implications which we are going to discuss in the next section. But before doing so, let us point out a possible caveat in our discussion. As we already mentioned, if the tunneling between the two vacua should go through the region where ${\cal Z}$ changes the sign, then $\sigma = {2\over \sqrt 3} (|V_{1}|^{1/2} + |V_{0}|^{1/2})$. In this case, if we ignore the decrease of the tension during the uplifting, the tunneling between the two vacua does not occur. One can clearly see it by looking at the gravitational suppression term $\bigl(1 -{9\sigma^{4}\over 16 |V_{1}|^{2}}\bigr)^{{-1}}$. This term for $\sigma = {2\over \sqrt 3} (|V_{1}|^{1/2} + |V_{0}|^{1/2})$ is negative, which corresponds to super-suppression of the tunneling \cite{Cvetic:1996vr}. 

On the other hand, we expect that in the realistic theories with hundreds of moduli fields  there should be many AdS minima, some of which will remain AdS after the uplifting. Therefore it is natural to expect that the decay channels at least to some of these vacua will be open, and will be described by  Eqs. (\ref{tunndsads}), (\ref{tunndsadsgrav}). This is the only thing that matters to us, since the total rate of the vacuum decay is determined by the path of the least resistance, i.e. by the channels which  maximize the tunneling probability.

\section{Sinks in the landscape  and the wave function of the universe}\label{sink}

\subsection{Tunneling, probabilities, and the wave function}

One of the main reasons to study vacuum transitions which occur on an incredibly large timescale is to find a probability distribution to live in one of the many vacuum states.
This issue has a long history involving quantum cosmology, eternal inflation, and statistical properties of the stringy landscape.

For a long time, vacuum transitions were studied in the context of inflationary cosmology.
The main subject was the possibility of transitions between dS vacuum and Minkowski vacuum, or between two dS vacua.
One of the most surprising and controversial results was obtained by Hawking and Moss \cite{Hawking:1981fz}.
They found that a phase transition from a dS minimum with energy density $V_{0}$ may occur via a quantum jump to the top of the barrier separating the two vacua, with the height $V_{1} > V_{0}$, and by a subsequent classical rolling to the second dS minimum with energy density $V_{3}$.
The probability of this jump, according to \cite{Hawking:1981fz}, was given by 
\begin{equation}
\label{HM} P = e^{-S_{1}+S_0}= \exp\left(-{24\pi^2\over V_0}+{24\pi^2\over V_1}\right) \ .
\end{equation}
Initial interpretation of this result was rather obscure because the corresponding instanton seemed to describe a homogeneous tunneling, which was certainly impossible in an infinite (or exponentially large) inflationary universe. Moreover, from the derivation of this result it was not clear why the tunneling should occur to the top of the potential instead of going directly to the second dS minimum.

A proper interpretation of the Hawking--Moss tunneling was achieved only after the development of the stochastic approach to inflation \cite{Starobinsky:1986fx,book,Linde:1991sk,LLM}.\footnote{For a Hamiltonian approach to the HM tunneling see \cite{Gen:1999gi}.} One may consider quantum fluctuations of a light scalar field $\phi$ with $m^2 = V'' \ll H^2 = V/3$.
During each time interval $\delta t = H^{-1}$ this scalar field experiences quantum jumps with the wavelength $\sim H^{-1}$ and with a typical amplitude $\delta\phi = H/2\pi$.
Then the wavelength of these fluctuations grows exponentially.
As a result, quantum fluctuations lead to a local change of the amplitude of the field $\phi$, which looks homogeneous on the horizon scale $H^{-1}$.
From the point of view of a local observer, this process looks like a Brownian motion of the homogeneous scalar field.
If the potential has a dS minimum at $\phi_0$ with $m\ll H$, then eventually the probability distribution to find the field with the value $\phi$ at a given point becomes time-independent, 
\begin{equation}
\label{E38a} P(\phi) \sim \exp\left(-{24\pi^2\over V_0}+{24\pi^2\over V(\phi)}\right) \ .
\end{equation}
That is why the probability to gradually climb to the local maximum of the potential at $\phi = \phi_{1}$ and then fall to another dS minimum is given by Eq. (\ref{HM}) \cite{Starobinsky:1986fx,book,Linde:1991sk,LLM}.

Note that the distribution $P(\phi)$ is a fraction of the {\it comoving} volume of the universe corresponding to each of the dS vacua. This probability distribution does not take into account different rate of growth of different part of the universe. Its interpretation can be understood as follows: At some initial moment one divides the universe into many domains of the same size, assigns one point to each domain, and follows the subsequent distribution  $P(\phi)$ of the points where  the scalar  field takes the value $\phi$.

A necessary condition for the derivation of this result in  \cite{Starobinsky:1986fx,book,Linde:1991sk,LLM} was the requirement that $m^2 = V'' \ll H^2 = V/3$.
This requirement is violated for all known scalar fields at the present (post-inflationary) stage of the evolution of the universe.
However, one may try to generalize it for the situations with $m^2 \gg H^2$.
Following \cite{Lee:1987qc} (see also \cite{Garriga:1997ef,Dyson:2002pf,Susskind:2003kw}), we will look for a probability distribution $P_{i}$ to find a given {\it point} in a state with the vacuum energy $V_{i}$. 
The main idea is to consider CDL tunneling between two dS vacua, with vacuum energies $V_{0}$ and $V_{1}$, such that $V_{0} < V_{1}$, and to study the possibility to tunneling in both directions, from $V_{0}$ to $V_{1}$, or vice versa.

The action on the tunneling trajectory, $S(\phi)$, does not depend on the direction in which the tunneling occurs, but the tunneling probability does depend on it.
It is given by $e^{-S(\phi)+S_0}$ on the way up, and by $e^{-S(\phi)+{ S_1}}$ on the way down  \cite{Lee:1987qc}. Let us  assume that the universe is in a stationary state, such that the comoving volume of the parts of the universe going upwards is balanced by the comoving volume of the parts going down.
This can be expressed by the detailed balance equation 
\begin{equation}
\label{balance} P_{0}\, e^{-S(\phi)+S_0} = P_{1}\, e^{-S(\phi)+S_1} \ , 
\end{equation}
which yields (compare with Eq.~(\ref{HM})) 
\begin{equation}
\label{weinb} {P_{1}\over P_{0}} = e^{-S_{1}+S_0} = \exp\left(-{24\pi^2\over V_0}+{24\pi^2\over V_1}\right) \ , 
\end{equation}
independently of the tunneling action $S(\phi)$.\footnote{Some problems arising in the derivation of this result in the Coleman-De Luccia approach will be discussed in \cite{lindeprep}.}

Equations (\ref{E38a}) and (\ref{weinb}) imply that the fraction of the comoving volume of the universe in a state $\phi$ (or $\phi_{1}$) different from the ground state $\phi_{0}$ (which is the state with the lowest, but positive, vacuum energy density) is proportional to $C_{0}\, \exp\bigl({24\pi^2\over V(\phi)}\bigr)$, with the normalization coefficient $C_{0} = \exp\bigl(-{24\pi^2\over V_{0}}\bigr)$. The probability distribution $C_{0}\, \exp\left({24\pi^2\over V(\phi)}\right)$ coincides with the square of the Hartle--Hawking wave function describing the ground state of the universe \cite{Hartle:1983ai}. It has a simple physical meaning: The universe wants to be in the ground state $\phi_{0}$ with the lowest possible value of $V(\phi)$, and the probability of the deviations from the ground state is exponentially suppressed.
This probability distribution also has a nice thermodynamic interpretation in terms of dS entropy  ${\bf S}$ \cite{Linde:1998gs}: 
\begin{equation}
\label{weinb2} {P_1\over P_{0}} = e^{{\bf S_{1}}-{\bf S_{0}}} = e^{{\bf \Delta S}}\ .
\end{equation}
Here, as before, ${\bf S_{i}} = - S_{i}$. This result and its thermodynamic interpretation played a substantial role in the discussion of the string theory landscape \cite{Susskind:2003kw}.

Unfortunately, there are some problems associated with this result.
It suggests that the universe is similar to a pond filled with still water, where all things accumulate at the bottom: The fraction of the comoving volume with the smallest possible vacuum energy must be overwhelmingly large, $P \sim \exp\left({24\pi^2\over V}\right)$.
For a discussion of paradoxes associated with this conclusion and their possible resolution in the KKLT scenario see, e.g.
\cite{Dyson:2002pf,Goheer:2002vf,Susskind:2003kw}.

Investigation of the stationary probability distribution alone does not give us a full picture. For example, the probability distribution (\ref{E38a}) tells us about  the fraction of the comoving volume of the universe in a given state, but it tells us nothing about the evolution towards this state. A partial answer to this question can be given by investigation of the stochastic diffusion equations describing the evolution of the scalar field in the inflationary universe. But now, instead of looking for the most probable outcome of the evolution, one should follow the evolution backwards and look for the initial condition $\phi_{0}$ for the trajectories which bring the field to its final destination $\phi$. In the stationary regime considered above, the corresponding solution looks very similar to (\ref{E38a}) \cite{LLM}:
\begin{equation}
\label{E38aa} P(\phi) \sim \exp\left(-{24\pi^2\over V(\phi_{0})}+{24\pi^2\over V(\phi)}\right) \ .
\end{equation}
In this equation, however, $\phi_{0}$ is not the position of the ground state, but a position of an arbitrary initial point for the diffusion process which eventually brings us to the point $\phi$. As we see, the probability is maximized by the largest possible value of $V(\phi_{0})$. Interestingly, expression $ \exp\bigl(-{24\pi^2\over V(\phi_{0})}\bigr)$ describing the probability of initial conditions coincides with the expression for the  square of the tunneling wave function describing creation of a closed dS universe ``from nothing'' \cite{Linde:1983mx}, whereas the second term looks like the square of the Hartle-Hawking wave function describing the ground state of the universe. In the stationary regime the squares of these two wave functions coexist in the same equation, but they provide answers to different questions.

Here we should note that it is far from being clear whether the simplest probability distribution which we studied so far should be used in anthropic considerations. 
 Indeed, the probability that something happen at a given {\it point}, which is described by the probability distribution in the {\it comoving volume} as discussed above, does not take into account the exponential growth of the {\it physical volume} of the universe, which is the main property of inflationary models.
For the discussion of the volume-weighted probability measure, see e.g. \cite{LLM,Bellido,Vilenkin:2006qf,LMprob,Tegmark}.

Despite this reservation,  it is quite important to study the probability distribution in the comoving volume since  its investigation is an essential step in several different attempts to calculate the probabilities in the landscape. 

Here we would like to discuss another problem with the stationary probability distribution described above: This stationary distribution is not very useful during inflation when one can ignore the existence of the lowest dS space with $V_{0} \sim 10^{{-120}}$; in order to obtain a stationary distribution one should take into account the growth of the physical volume of the universe   \cite{LLM}.  Moreover, as we are going to show now,  this distribution does not necessarily apply to the string theory landscape, simply because in the KKLT scenario there are no stable dS vacua that could serve as a ground state of the universe.
Metastability of dS space in the KKLT construction was emphasized in \cite{Kachru:2003aw} and in many subsequent papers.
Here we would like to look at this issue in a more detailed way.

\subsection{Probability currents and sinks in the landscape}

Stationarity of the probability distribution (\ref{weinb2}) was achieved because the lowest dS state did not have any further way to fall.
Meanwhile, in string theory it is always possible for the metastable dS state to decay.
It is important that if it decays by production of the bubbles of 10D Minkowski space \cite{Kachru:2003aw}, or by production of bubbles containing a collapsing open universe with a negative cosmological constant \cite{Coleman:1980aw}, the standard mechanism of return back to the original dS state does not operate any longer.\footnote{One may speculate about the possibility of quantum jumps from Minkowski space to dS space \cite{Linde:1991sk}, or even about the possibility of jumps back through the cosmological singularity inside each of the bubbles, but we will not discuss these options here.} These processes work like sinks for the flow of probability in the landscape. The fraction of the comoving volume in dS vacua will decrease in time.

To get a visual understanding of this process, assume that we are looking at the universe in the comoving coordinates, and paint black all of its parts corresponding to one of the two dS states, and paint white the parts in another dS state.
Then the stationary regime which we discussed so far would imply that eventually the whole universe in average becomes gray, and the level of gray asymptotically will remain constant.

Suppose now that some parts of the universe may tunnel to a state with a negative cosmological constant.
These parts will collapse, so they will not return to the initial dS vacua.
If we paint such parts red, then the universe, instead of reaching the constant gray level, eventually will look completely red.

This does not mean that eventually all parts of our universe will end up in the sink.
Due to the exponential expansion of the universe in various dS vacua, the universe eternally rejuvenates, and the total volume of the universe in different dS states continues to grow exponentially \cite{linde1982,Guth:1980zm,Steinhardt,Vilenkin:1983xq}.
However, if one wants to find the ratio of the comoving volume of the universe in different states, then instead of solving the detailed balance equation (\ref{balance}), one should solve the ``vacuum dynamics'' equations 
\begin{equation}
\label{v01} \dot P_{0} = - J_{0s} -J_{01} + J_{10} \ , 
\end{equation}
\begin{equation}
\label{v02} \dot P_{1} = - J_{1s} -J_{10}+J_{01} \ .
\end{equation}
Here $ J_{0s} = P_{0}\,e^{-C_{0}}$ is the probability current from the lower dS vacuum to the sink,  i.e. to a collapsing universe, or to a Minkowski vacuum, $ J_{1s} = P_{1}\, e^{-C_{1}}$  is the probability current from the upper dS vacuum to the sink, $ J_{01} =P_{0}\, e^{-{\bf S_{0}}+|S(\phi)|}$ is the probability current from the lower dS vacuum to the upper dS vacuum, and $ J_{10} = P_{1}\, e^{-{\bf S_{1}}+|S(\phi)|}$ is the probability current from the upper dS vacuum to the lower dS vacuum. Combining this all together, gives us the following set of equations for the probability distributions: 
\begin{equation}
\label{v1} \dot P_{0} = - P_{0}\, e^{-C_{0}} -P_{0}\, e^{-{\bf S_{0}}+|S(\phi)|} + P_{1}\, e^{-{\bf S_{1}}+|S(\phi)|} \ , 
\end{equation}
\begin{equation}
\label{v2} \dot P_{1} = - P_{1}\, e^{-C_{1}} -P_{1}\, e^{-{\bf S_{1}}+|S(\phi)|} + P_{0}\, e^{-{\bf S_{0}}+|S(\phi)|} \ .
\end{equation}
Since the entropy is inversely proportional to the energy density, the entropy of the lower level is higher, ${\bf S_{0}} > {\bf S_{1}}$. Since the tunneling is exponentially suppressed,  we have ${\bf S_{1}} > |S(\phi)|$, so we have a
hierarchy $ {\bf S_{0}} > {\bf S_{1}} > |S(\phi)|$. We will associate the lower vacuum with our present vacuum state, with $S_{0} \sim 10^{{120}}$, and therefore  in what follows  we will often have in mind the regime $ {\bf S_{0}} \gg {\bf S_{1}} > |S(\phi)|$.

For simplicity, we will study here the  possibility that only the lower vacuum can tunnel to the sink, i.e.
we will take the limit $C_{1}\to \infty$ and drop the term $- J_{1s}= - P_{1}\, e^{-C_{1}}$ in Eq. (\ref{v2}). We  will analyze more general solutions of equations (\ref{v1}), (\ref{v2}) in a separate publication \cite{lindeprep}.
On the other hand, we will keep in mind the results of the previous Section, where we have found that typically the probability of the decay of a  metastable dS vacuum to a sink  can be quite high, $ e^{-C_{0}} \sim \exp \bigl(-{O(m_{3/2}^{-2})}\bigr) \gg e^{-{\bf S_{0}}} \sim e^{-10^{120}}$.  Therefore we  expect that $C_{0} \ll {\bf S_{0}}$.
Other mechanisms of a relatively fast vacuum decay have been discussed e.g.
in \cite{Frey:2003dm,Green:2006nv,Danielsson:2006jg}.
The main conclusions to be obtained in this section will be valid for any of these mechanisms.

By solving equations (\ref{v1}), (\ref{v2}), one can show that $P_{0}$ and $P_{1}$ exponentially decrease in the course of time due to the existence of the sink (the term $-J_{0s} =- P_{0}\, e^{-C_{0}}$), but the corresponding physical volume of the universe exponentially grows, and the ratio $P_{1}(t)/P_{0}(t)$ approaches a stationary regime $P_{1}(t)/P_{0}(t) = p = const$.
In order to find $p$, one can add to each other our equations (without the term $ - P_{1}\, e^{-C_{1}}$). This yields 
\begin{equation}
\label{v3} (1+p)\dot P_{0} = - P_{0}\, e^{-C_{0}} \ .
\end{equation}
The solution is 
\begin{equation}
P_{0} = {P_{1}\over p} = - P_{0}(t=0)\ \exp\left(-{e^{-C_{0}}\over 1+p}\, t\right) \ .
\end{equation}

As for  the (asymptotically) constant ratio  $p  = P_{1}(t)/P_{0}(t)$, from eqs.~(\ref{v1}), (\ref{v3}) one finds 
\begin{equation}
\label{ttt} 1 = (1+p)\left(1+e^{C_{0}-{\bf S_{0}}+|S(\phi)|} - p\,e^{C_{0}-{\bf S_{1}}+|S(\phi)|}\right) \ .
\end{equation}

One may consider two interesting regimes. Suppose first that  $ e^{-C_{0}} \ll e^{-{\bf S_{1}}+|S(\phi)|}$, i.e. the probability to fall to the sink from the lower vacuum is smaller than the probability of the decay of the upper vacuum.
In this case one recovers the previous result, Eq.~(\ref{weinb2}), which is related to the square of the Hartle-Hawking wave function: 
\begin{equation}
\label{HHeq} p = {P_{1}\over P_{0}} = e^{{\bf S_{1}}-{\bf S_{0}}} \ll 1 \ .
\end{equation}

Now let us consider an opposite regime, and assume that the decay rate of the uplifted dS vacuum to the sink is relatively large, $ e^{-C_{0}} \gg e^{-{\bf S_{1}}+|S(\phi)|}$ (which automatically means that   $ e^{-C_{0}} \gg e^{-{\bf S_{0}}+|S(\phi)|}$). In this case the solution of Eq.~(\ref{ttt}) is 
\begin{equation}\label{ooo}
p = {P_{1}\over P_{0}} = e^{{\bf S_{1}}-|S(\phi)| -C_{0}} \approx e^{{\bf S_{1}}-|S(\phi)|} \gg 1 \ , 
\end{equation}
i.e. one has an inverted probability distribution. This result has a simple interpretation: if the ``thermal exchange'' between the two dS vacua occurs very slowly as compared to the rate of the decay of the lower dS vacuum, then the main fraction of the volume of the dS vacua will be in the state with higher energy density.

Note that the total fraction of the comoving volume remaining in both of the dS spaces exponentially decreases in time due to the existence of the sink in the lower dS vacuum:
\begin{equation}\label{sinkpower}
P_{0}+P_{1} \sim  \exp\left(-{e^{-C_{0}}\over1+p}\, t\right) \ .
\end{equation}
This result  may have  interesting methodological implications.
In many recent discussions of the probabilities in the landscape the authors consider the world as seen by an imaginary eternal observer.
In the context of the stringy landscape  such considerations are problematic, because every would-be eternal observer living in the landscape (unlike an observer in an eternally existing dS space) begins his life in a cosmological singularity and ends up in a sink, within the typical time determined by Eq. (\ref{sinkpower}). Moreover, if the sinks occur in the higher dS vacua (see below), then the ``eternal observers'', or markers, will have a good chance to die well before reaching the thermalized regions produced after inflation.  Introduction of a finite lifetime of such observers may result in the modification of the probability measure as suggested in \cite{LMprob}.

Here we should make an important clarification. The probability distribution $P_{i}$ describes the fraction of the volume of the universe in a particular dS state. However, when the bubbles of a new phase expand, their interior becomes an empty dS space. If we are  usual observers which are born after reheating of the universe, then one may argue that the probability {\it to live} in the  bubble $dS_{i}$ is proportional not to the relative volume of the bubbles $p = P_{1}/P_{0}$ but to the frequency of their production, i.e. to the ratio of the probability currents $J_{01}/J_{10}$ \cite{Bellido,Vilenkin:2006qf}. In the absence of the sink,  the fraction of the comoving volume which flows to the lower dS vacuum due to the tunneling from the upper dS vacuum is equal to the fraction of the volume jumping upwards from the lowest vacuum to the higher vacuum. In other words, the two probability currents are exactly equal to each other, 
\be
J_{01} = J_{10}  \ , 
\ee
which is the essence of the detailed balance equation (\ref{balance}). Interestingly, our new results imply that this regime remains approximately valid even in the presence of the sink, under the condition $e^{-C_{0}} \ll e^{-{\bf S_{1}}+|S(\phi)|}$.
 
On the other hand, in the regime described by Eq.  (\ref{ooo}), which occurs if the decay rate to the sink is large enough, $e^{-C_{0}} \gg e^{-{\bf S_{1}}+|S(\phi)|}$, one has a completely different result:
 \begin{equation}
\label{vvvv} {J_{01}\over J_{10}} =  {P_{0}\, e^{-{\bf S_{0}}+|S(\phi)|}\over  P_{1}\, e^{-{\bf S_{1}}+|S(\phi)|}} =  e^{-{\bf S_{0}}+|S(\phi)| +C_{0}} \approx e^{-{\bf S_{0}}} \sim e^{-10^{120}} \ . 
\end{equation}
Thus we have a crucial regime change at the moment when the decay rate of the lower vacuum to the sink starts competing with the decay rate of the upper dS vacuum.

Now we will compare the probability current  to the sink, $J_{0s}$, and the probability current from the lower dS upwards, $J_{01}$. In the absence of the sink, $J_{0s}$ was zero, so every point falling to the lower dS vacuum eventually jumps up to the higher dS vacua. This leads to eternal recycling of dS vacua \cite{Hawking:1981fz,Lee:1987qc,Garriga:1997ef}, and to the problems discussed in \cite{Dyson:2002pf,Goheer:2002vf}: In the absence of the sinks (i.e. in the models with eternal dS space) each point in the comoving volume would indefinitely wonder between different dS spaces. In this scenario, it would be more probable that our part of the universe was formed as a result of a large quantum fluctuation upwards from an empty dS space, instead of being formed at the stage of inflation. This would leave the homogeneity of our universe, and the origin of perturbations of metric unexplained   \cite{Dyson:2002pf,Goheer:2002vf}. 

Invention of the KKLT mechanism demonstrated that all dS vacua in string theory are metastable \cite{Kachru:2003aw}, with the lifetime smaller than the recurrence time $t_{r}  \sim  e^{\bf S_{0}} \sim e^{10^{120}}$. This means that we do not have much time  for  the  incredibly improbable jumps upwards to happen \cite{Goheer:2002vf}. However, the time necessary for the improbable fluctuations upwards, which could re-create conditions for the existence of life, is also somewhat smaller than the recurrence time \cite{Goheer:2002vf}. Therefore in order to make sure that the universe does not recycle back from our dS state we would need to double-check that the probability to jump upwards is indeed strongly suppressed as compared to the probability to tunnel to the sink. 

The estimates of the decay probability of dS vacuum contained in our paper confirm that this is indeed the case.  Indeed,  investigation of the tunneling in several models discussed in our paper suggests that the probability that any given part of dS space with $V \sim 10^{-120}$ will jump upwards and participate in the process of recycling of the universe is  suppressed by a factor of $O(e^{-10^{120}})$. 
This can be expressed in terms of the ratio of the probability current from the lower dS vacuum to the upper vacuum $J_{01}$ with the probability current from the lower dS vacuum to the sink $J_{0s}$:
 \begin{equation}
\label{vvvvv} {J_{01}\over J_{0s}} =  e^{-{\bf S_{0}}+|S(\phi)| +C_{0}} \approx  e^{-{\bf S_{0}}}  \sim e^{-10^{120}} \ . 
\end{equation}
The last part of this equation follows from  the condition  $ {\bf S_{0}} \gg {\bf S_{1}} > |S(\phi)|$, and from the estimate of the decay rate obtained in the previous Section, which suggests that in many cases this rate is quite high, $C_{0} \ll {\bf S_{0}}$.

 Despite this fact, recycling may still be possible, since for $e^{-C_{0}} \ll e^{-{\bf S_{1}}+|S(\phi)|}$ one has $J_{01} = J_{10}$. On the other hand, for $e^{-C_{0}} \gg e^{-{\bf S_{1}}+|S(\phi)|}$ one has $J_{01} = e^{-10^{120}} J_{10}$. This suggests that in this regime the probability that our universe was created due to the jumps up \cite{Dyson:2002pf,Goheer:2002vf} becomes exponentially small as compared with the probability that it was formed as a result of inflation. Using our results, one can easily check that this is indeed the case if the rate of spontaneous formation of the parts of the universe of our type in the vacuum $dS_{0}$ is smaller than the decay rate of the vacuum $dS_{0}$ to the sink. For wide sinks, this condition can be easily satisfied.

This conclusion is closely related to the inversion of the probability distribution for $e^{-C_{0}} \gg e^{-{\bf S_{1}}+|S(\phi)|}$ (wide sink regime), see Eq. (\ref{ooo}).  Note that inverted probability distributions in the presence of the sink represented by  Minkowski vacuum naturally appear in the simplest chaotic inflation scenario with the volume-weighted probability measures proposed in \cite{LLM,Bellido}; see Figs. 10 and 12 in \cite{LLM}, where the basic theory of the volume-weighted distributions in the inflationary cosmology was developed.  With the first of the two volume-weighted probability measures proposed in \cite{LLM,Bellido} the problem of non-inflationary jumps up does not appear at all; the growth of the total volume of the universe (and the corresponding fluxes) is predominantly determined by the exponential expansion of domains with larger vacuum energy and by the subsequent  inflation on the way down.

In our investigation so far we considered the simplest model with two dS vacua, with the probability leak in the lower dS vacuum.
On the other hand, if the probability leak occurs only in the upper dS vacuum, one may ignore the term $- J_{0s} = - P_{0} e^{-C_{0}}$ in Eq.~(\ref{v1}) but keep the term $ - J_{1s} = -P_{1} e^{-C_{1}}$ in Eq.~(\ref{v2}). In this case, for  $ e^{-C_{1}} \ll e^{-{\bf S_{1}}+|S(\phi)|}$    one again recovers the distribution (\ref{HHeq}), whereas for $ e^{-C_{1}} \gg e^{-{\bf S_{1}}+|S(\phi)|}$ one finds
\begin{equation}
p = {P_{1}\over P_{0}} = e^{{\bf -S_{0}}+|S(\phi)|+C_{1}} \approx e^{{\bf -S_{0}}}\sim  e^{-10^{120}} \ . 
\end{equation}
Note that in this regime we have a flat probability distribution, which does not depend on $V(\phi_{1})$.

More generally, one should consider a dynamical equilibrium of a system of many dS, Minkowski and AdS vacua, where the probability leaks may occur at each level.
The fraction of the comoving volume occupied by different metastable vacua in such a universe will be exponentially sensitive to the decay modes for each of these vacua. This makes the investigation of the probabilities in the landscape much more involved, but also much more interesting. Previously we were trying to construct the wave function of the ground state of the universe. Now we should learn how to study the universe with the holes in the ground.

\section{Discussion}

There are two possible approaches to string theory landscape. The first one is based on the calculation of the wave function of the universe, which may describe all branches of the wave function corresponding to all possible vacua. In fact, the first two proposals for the anthropic solution of the cosmological problem in the context of inflationary cosmology were based on the investigation of quantum creation of the universe with various values of scalar fields and fluxes \cite{Linde:1984ir}, or with different types of compactification \cite{Sakharov}.

Another approach is to consider an inflationary universe consisting of many parts corresponding to all possible vacua \cite{linde1982}, which allows, in particular, to consider a single universe divided into parts corresponding to all possible values of  scalar fields,  fluxes and compactifications, and look for the anthropic solution of the cosmological constant problem in this context, see e.g.  \cite{Banks,300,Weinberg:1987dv,Brown:1988kg,Bousso:2000xa,Feng:2000if}. Conceptually, this approach is much simpler than the approach based on quantum cosmology, but in order to make it sufficiently general one must find a way to describe the universe consisting of many different parts and describe possible transitions between them. This was the goal of our paper.

One of the problems here  is that  a fully consistent ``landscape action'' with the bulk and brane sources is not known  \cite{Banks:2003es,Denef,deAlwis:2006cb}.  However,  one can hope that such action can be constructed at least in the case of unbroken supersymmetry following the ``supersymmetry in singular spaces'' proposal \cite{Bergshoeff:2000zn}.  This strategy was already applied in examples of  fully consistent supersymmetric bulk and brane actions in \cite{Bergshoeff:2001pv} which include the D8 domain walls with the piece-wise constant $G_0$ flux.   In a more general situation with moduli stabilization when various fluxes may jump due to brane sources,  the corresponding construction has still to be developed.  

Therefore in the current project we restricted ourselves  to studies of the regions of the landscape where stabilization of the moduli is under control and  fluxes are fixed. Even in this restricted and simplified situation we were able to find multiple vacua and BPS domain walls separating them. 

In the first part of this paper we concentrated on the stringy landscape prior to the uplifting and found many BPS domain wall solutions separating different parts of the stringy landscape from each other. We did it using the methods developed earlier in \cite{Cvetic:1996vr}, and also more recent methods based on the ``new attractor equations'' \cite{Kallosh:2005ax}.

In the second part of the paper we studied what  happens with the domain walls after the uplifting. In general, potentials of many moduli before and after uplifting may have many different AdS minima of the moduli potential, for the same values of fluxes. Before the uplifting, the tunneling between these minima was forbidden by supersymmetry. One could expect, therefore, that after the uplifting the rate of tunneling may be related to the degree of supersymmetry breaking. Indeed,  in the particular examples that we have studied the probability of the decay of the uplifted vacuum to the remaining AdS vacuum was suppressed by a factor $\sim \exp \bigl(-{O(m_{3/2}^{-2})}\bigr)$, where $m_{{3/2}}$ is the gravitino mass. This decay probability is much greater than $e^{-{\bf S}} \sim e^{-10^{120}}$.

One of the interesting features of the decay of  dS vacua to Minkowski or AdS vacua is that this decay is irreversible. This process is substantially different from the stationary ``thermal exchange''  between various dS vacua. The existence of the channels of the irreversible vacuum decay, which serve as sinks for the flow of probability, is a distinguishing new feature of the stringy landscape which deserves further investigation. As we show in Section 6 of our paper, the existence of the sinks strongly affects the probability distributions in string cosmology. 

\ 

\noindent{\large{\bf Acknowledgments}}

\noindent
It is a pleasure to thank L. Alvarez--Gaum\'e, R. Bousso, T. Banks, M.~Cveti\v{c}, F. Denef,  B. Freivogel, G. Horowitz, S. Kachru, M. Mari\~no, D. Martelli,  E. Silverstein, S. Shenker,  L. Susskind and W. Taylor for stimulating conversations.  The work of A.C. is supported in part by the European Community's Marie Curie Actions, Project No. MRTN-CT-2004-005104 (Forces Universe). The work of R.K., A.L. and A.G. was supported by NSF grant PHY-0244728. The work of A.G. was also supported by NSF grant PHY-0097915, and by DOE under contracts DE-AC03-76SF00515 and RFBR 05-01-00758.



\vskip 2cm

\providecommand{\href}[2]{#2}\end{document}